\documentclass[conference]{IEEEtran}
\IEEEoverridecommandlockouts
\usepackage{cite}
\usepackage{amsmath,amssymb,amsfonts}
\usepackage{algorithmic}
\usepackage{graphicx}
\usepackage{textcomp}
\usepackage{xcolor}
\def\BibTeX{{\rm B\kern-.05em{\sc i\kern-.025em b}\kern-.08em
    T\kern-.1667em\lower.7ex\hbox{E}\kern-.125emX}}

\graphicspath{{figures/}}
\usepackage[tight,footnotesize]{subfigure}

\begin{document}

\title{ResumeNet: A Learning-based Framework for Automatic Resume Quality Assessment
}


\author{\IEEEauthorblockN{Yong Luo$\ast$}
\IEEEauthorblockA{\textit{School of Computer Science and Engineering} \\
\textit{Nanyang Technological University}\\
Singapore \\
yluo180@gmail.com}
\and
\IEEEauthorblockN{Huaizheng Zhang$\ast$}
\IEEEauthorblockA{\textit{School of Computer Science and Engineering} \\
\textit{Nanyang Technological University}\\
Singapore \\
huaizhen001@e.ntu.edu.sg}
\and
\IEEEauthorblockN{Yongjie Wang}
\IEEEauthorblockA{\textit{School of Computer Science and Engineering} \\
\textit{Nanyang Technological University}\\
Singapore \\
n1705223J@e.ntu.edu.sg}
\and
\IEEEauthorblockN{Yonggang Wen}
\IEEEauthorblockA{\textit{School of Computer Science and Engineering} \\
\textit{Nanyang Technological University}\\
Singapore \\
ygwen@ntu.edu.sg}
\and
\IEEEauthorblockN{Xinwen Zhang}
\IEEEauthorblockA{\textit{Hiretual} \\
\textit{Mountain View, California}\\
USA \\
xinwenzhang@hiretual.com}
\thanks{$^\ast$Yong Luo and Huaizheng Zhang contributed equally to this work.}
}


\maketitle

\begin{abstract}
Recruitment of appropriate people for certain positions is critical for any companies or organizations. Manually screening to select appropriate candidates from large amounts of resumes can be exhausted and time-consuming. However, there is no public tool that can be directly used for automatic resume quality assessment (RQA). This motivates us to develop a method for automatic RQA. Since there is also no public dataset for model training and evaluation, we build a dataset for RQA by collecting around $10K$ resumes, which are provided by a private resume management company. By investigating the dataset, we identify some factors or features that could be useful to discriminate good resumes from bad ones, e.g., the consistency between different parts of a resume. Then a neural-network model is designed to predict the quality of each resume, where some text processing techniques are incorporated. To deal with the label deficiency issue in the dataset, we propose several variants of the model by either utilizing the pair/triplet-based loss, or introducing some semi-supervised learning technique to make use of the abundant unlabeled data. Both the presented baseline model and its variants are general and easy to implement. Various popular criteria including the receiver operating characteristic (ROC) curve, F-measure and ranking-based average precision (AP) are adopted for model evaluation. We compare the different variants with our baseline model. Since there is no public algorithm for RQA, we further compare our results with those obtained from a website that can score a resume. Experimental results in terms of different criteria demonstrate effectiveness of the proposed method. We foresee that our approach would transform the way of future human resources management.
\end{abstract}


\section{Introduction}
\label{sec:Introduction}
It is critical to recruit appropriate talents for certain positions in either companies or organizations. A company, especially the big ones such as ``Google'', often receives hundreds of thousands of resumes for job application every year. Besides, headhunting or human resource (HR) managers may search for talents on job service platform, such as ``LinkedIn'' and ``Monster'', to find appropriate candidates for certain job positions. The searched resulting list of resumes may be very long. Manually screening all the resumes to select possible candidates for further consideration, such as interview, is labor-intensive and time-consuming. From the talent perspective, he/she may want to know whether his/her resume is good enough. Therefore, it is desirable to develop some tool to assess the quality of each resume automatically.

Although there exist some resume quality assessment (RQA) websites (e.g., http://rezscore.com/), their underlying assessment schemes or algorithms are unknown and there is no public dataset for model training and evaluation. To tackle these issues, we build a dataset and develop a general model to assess the quality of resumes automatically.

The dataset is built by collecting resumes from a private resume management company. We collect around $10,000$ resumes in total, and some of them are labeled by experts (e.g., the HR managers). Each label indicates whether an expert is interested in a resume or not. The number of labeled resumes is very small since most of the resumes are not reviewed and only a few experts would like to give feedback. We preprocess the resumes and extract some information that may be useful to determine the quality of a resume, e.g., the consistency between different parts of a resume, the user education level, skills and working experiences. Then we develop a neural-network framework to predict a score for each resume by integrating the different types of information. In the framework, some text processing techniques, such as word and sentence embedding \cite{D-Cer-et-al-arXiv-2018} are incorporated. We also introduce the attention scheme \cite{JL-Yang-et-al-CVPR-2017} to integrate multiple word/sentence embeddings.

To train a specific model under the framework, it is straightforward to adopt the least squared loss. However, this may lead to unsatisfactory results since the provided labeled data are limited. Besides, we found that the positive samples (good resumes) are much fewer than the negative samples in the dataset. Both of the label deficiency and class-imbalance issues are popular in real-world applications due to the high labeling cost \cite{X-Zhu-TR-Madison-2006, O-Chapelle-et-al-TNN-2009, A-Isabelle-et-al-NIPSw-2017, N-Japkowicz-and-S-Stephen-IDA-2002}. To deal with these issues, we propose two variants of the model by utilizing the pair and triplet based losses inspired by \cite{S-Chopra-et-al-CVPR-2005, M-Norouzi-et-al-NIPS-2012, F-Schroff-et-al-CVPR-2015}. These losses are designed for embeddings and we revise them so that they are appropriate for the prediction scores in our framework. We can generate large amounts of training pairs or triplets given only a few single labeled samples. Employing the pair/triplet-based loss is also advantageous in that the outputs of different positive/negative samples are not enforced to approach the same value ($1$ or $-1$). Hence, the within-class difference is respected. The hard label is relaxed to be soft label and the generalization ability of the model is improved. Since there are large amounts of unlabeled samples, we further propose an alternative strategy to alleviate the label deficiency issue by exploiting the structure of the data distribution. This is conducted by adding a manifold regularization (MR) term \cite{M-Belkin-et-al-JMLR-2006} to include the abundant unlabeled data in the training.

Overall, the main contributions of this paper are:
\begin{itemize}
  \item We build a novel dataset and identify some discriminative factors or features for automatic resume quality assessment (RQA).
  \item We develop a general framework for RQA. As far as we are concerned, this is the first public work for this application.
  \item We propose several strategies to handle the label deficiency issue, which is common in real-world applications.
\end{itemize}
We compare the baseline model with its different variants by varying the loss or making use of the unlabeled data. To evaluate the effectiveness of different models, we adopt three different criteria: receiver operating characteristic (ROC) curve, F-measure and average precision (AP). The ROC curve and F-measure are popular for evaluating the classification performance. AP is able to evaluate the performance of ranking according to the scores. This is useful when we would like to sort a list of resumes and select the top-rank ones. Since there is no public algorithm for RQA, we further compare our results with those obtained from a resume assessment website. We conduct extensive experiments and the results under various criteria demonstrate effectiveness of the proposed method.

The rest of the paper is organized as follows. In Section \ref{sec:Related_Work}, we first review some related work on text embedding and feature aggregation, and then summarize some techniques that deal with the label deficiency issue in machine learning. Section \ref{sec:Overview} is an overview of our dataset and framework for RQA. The objective functions of the baseline model and its variants are presented in Section \ref{sec:Formulation}. Section \ref{sec:Experiments} are some experimental results and finally we give some conclusions and insights in Section \ref{sec:Conclusion}.

\section{Related Work}
\label{sec:Related_Work}

\subsection{Text Embedding}
In the text analysis area, there is an increasing interest on the embedding technique \cite{H-Gui-et-al-ICDM-2016}. For example, word/phrase embedding \cite{Y-Bengio-et-al-JMLR-2003, GX-Xun-et-al-ICDM-2016} is to learn a compact continuous-valued vector to represent a word/phrase. This is superior to the traditional one-hot representation, which often suffers the curse-of-dimensionality and data sparsity problems. A summarization of some popular word embedding approaches can be found in \cite{J-Turian-et-al-ACL-2010}, and word2vec \cite{T-Mikolov-et-al-NIPS-2013} and Glove \cite{J-Pennington-et-al-EMNLP-2014} are two representative works.

Recently, some sentence embedding methods \cite{A-Conneau-et-al-EMNNLP-2017, D-Cer-et-al-arXiv-2018} are developed to map a sentence into a vector. This is advantageous in that the correlations among multiple words can be exploited. In this paper, we choose the universal sentence encoder (USE) \cite{D-Cer-et-al-arXiv-2018} to embed both the words and sentences since we found that satisfactory embeddings can be obtained for both of them. Besides, the resulting embeddings of words and sentences are comparable, and this will facilitate the subsequent learning in our framework.


\subsection{Feature Aggregation}
In either the textual and visual analytic-based applications, we often need to combine multiple representations to obtain a unified representation. For example, there are multiple words in a sentence and a video consists of multiple frames. To aggregate multiple feature representations, the most commonly utilized strategies may be average and max pooling \cite{B-Graham-arXiv-2014}. However, these strategies treat each representation equally and simply ignore the different importance of different representations. Multi-view learning \cite{Y-Luo-et-al-TKDE-2015, Y-Luo-et-al-TIP-2015, YJ-Fu-et-al-ICDM-2016} can learn a weight to reflect the importance (such as discriminative ability) for each type of feature, but it usually cannot handle the varied number of input features and is sensitive to the representation order.

To overcome these drawbacks, some recent works \cite{O-Vinyals-et-al-ICLR-2016, JL-Yang-et-al-CVPR-2017} propose to learn adaptive weight by applying the attention mechanism \cite{S-Poriaa-et-al-ICDM-2017}. The dimension of the learned attention parameter is the same as the representation and hence these approaches can handle arbitrary number of inputs and invariant to the input order.

\subsection{Learning with Limited Labeled Data}
It is common that we have only a few labeled data in real-world applications due to the high-labeling cost. There are two popular techniques that can be used to tackle the label deficiency issue: semi-supervised learning \cite{X-Zhu-TR-Madison-2006, O-Chapelle-et-al-TNN-2009, Y-Luo-et-al-TIP-2013} and transfer learning \cite{Y-Luo-et-al-TIP-2014, Y-Luo-et-al-TPAMI-2018}. The former assumes there are abundant unlabeled data to help the model training, while the latter assumes there are some different but related source domains, where abundant labeled data are available. In this paper, we choose the former since we have large amounts of unlabeled data and employ the well-known manifold regularization \cite{M-Belkin-et-al-JMLR-2006, Y-Luo-et-al-TNNLS-2013} method to improve the performance.

Alternatively, we can increase the number of labeled data by adopting the pair or triplet based loss \cite{S-Chopra-et-al-CVPR-2005, F-Schroff-et-al-CVPR-2015}. Such loss is usually used for fine-grained recognition \cite{D-Cheng-et-al-CVPR-2016} or search \cite{J-Wang-et-al-CVPR-2014} to discriminate objects that have minor differences. A byproduct is that large amounts of training pairs or triplets can be generated even if only a few labeled samples are provided. The Siamese network \cite{S-Chopra-et-al-CVPR-2005} and FaceNet \cite{F-Schroff-et-al-CVPR-2015} are two representative works. However, the outputs of these approaches are embeddings. While in our method, the outputs are prediction scores, which should be large (resp. small) for positive (resp. negative) samples. Such constraints are stronger than those for embeddings and hence the losses cannot be directly employed. In this paper, we revise them so that they are appropriate for our method.

\begin{figure}[htbp]
\centering
\includegraphics[width=1.0\columnwidth]{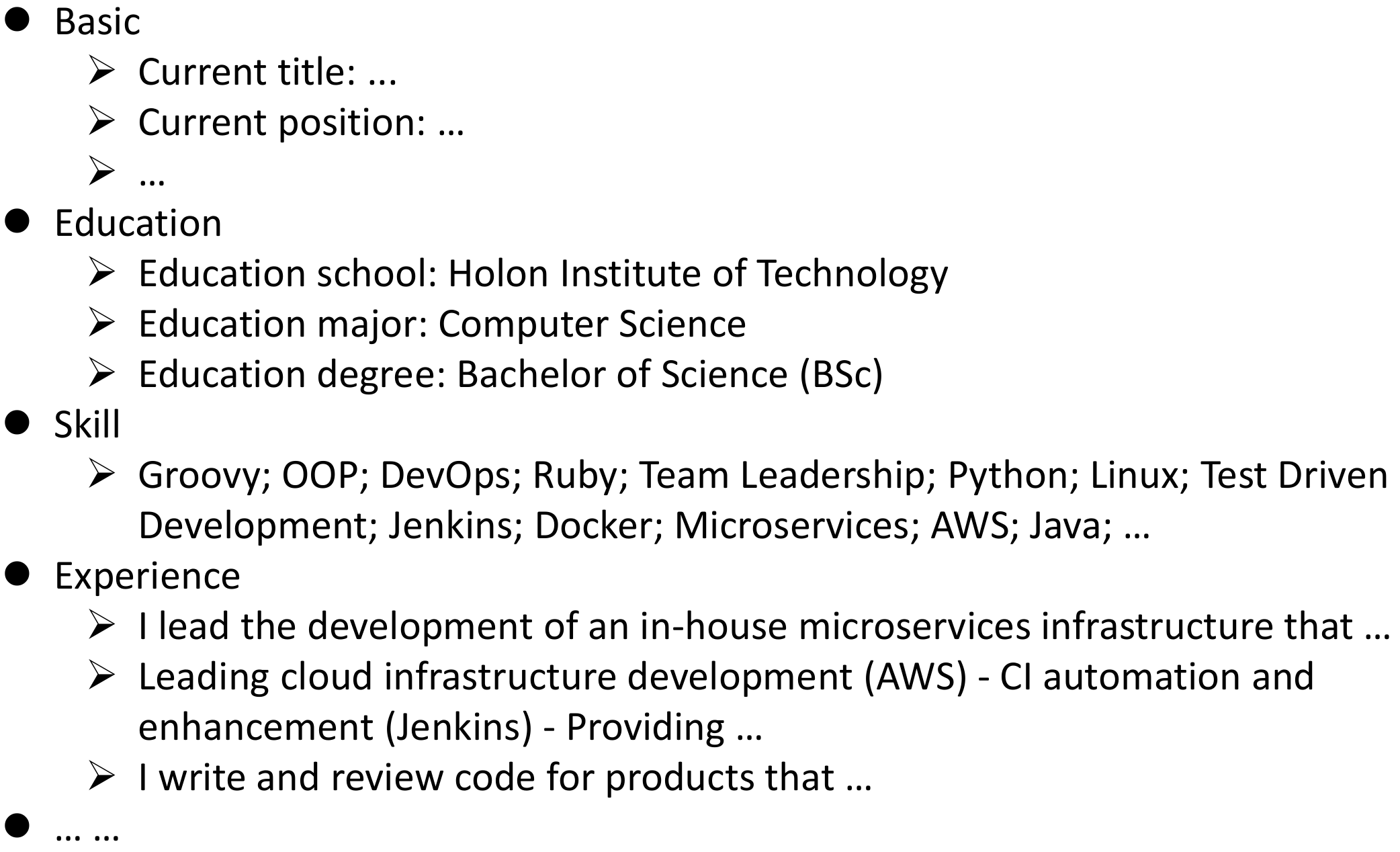}
\caption{An example of the information provided in a resume.}
\label{fig:Resume_Example}
\end{figure}

\begin{figure*}[htbp]
\centering
\includegraphics[width=1.3\columnwidth]{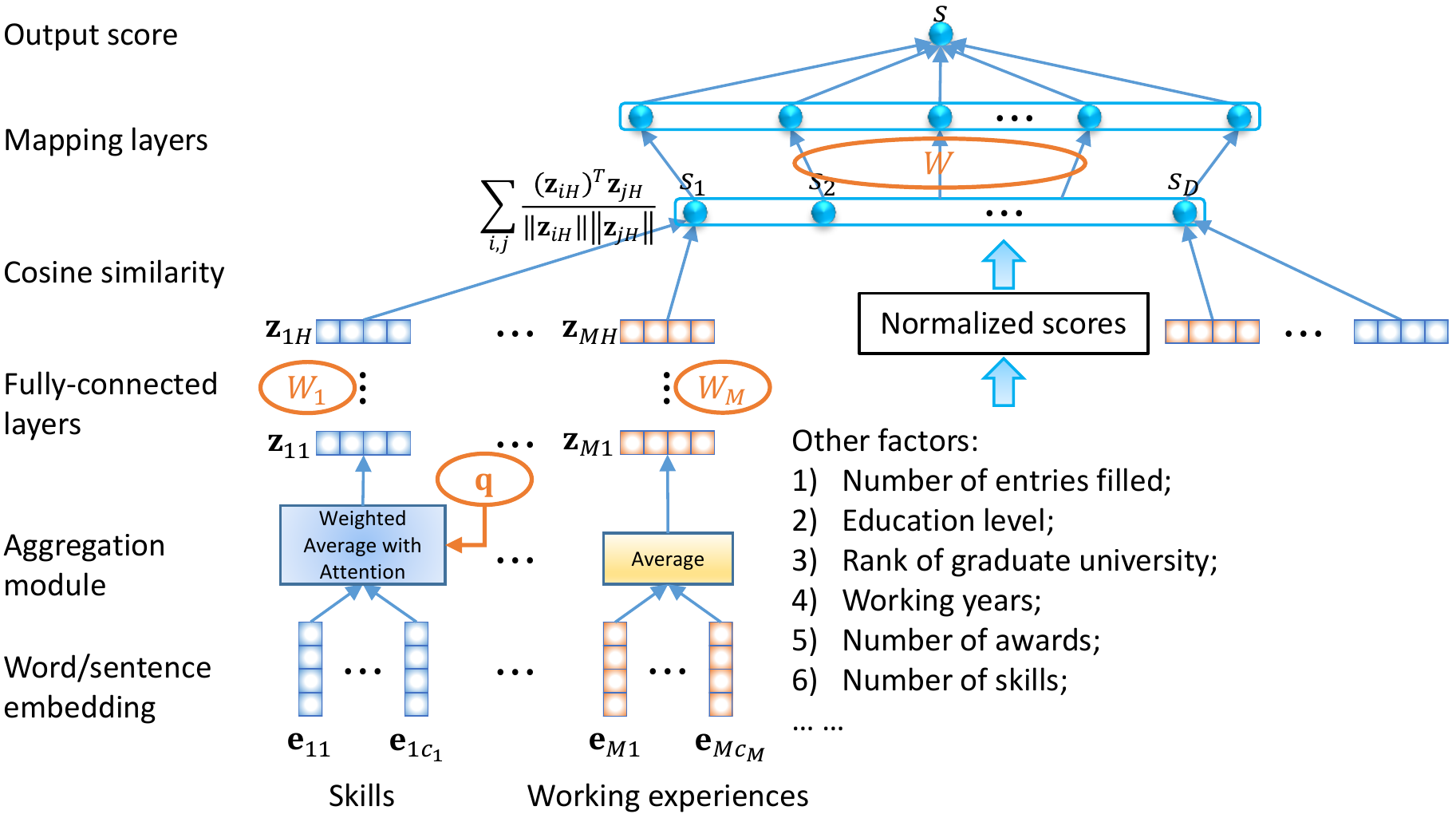}
\caption{Architecture of the proposed neural network framework for resume quality assessment. The framework mainly consists of a sub-network to measure the consistency of a resume's different parts, and some factors, which are mapped into continuous values. In the sub-network, the input words or sentences are first mapped into continuous-valued embeddings, and then simple average or weighted average using the attention scheme is applied to aggregate different embeddings as a single embedding. Some fully-connected layers can be added to transform the unified embeddings of different parts as comparable latent representations, and the cosine similarity of different representations is utilized to measure the consistency between different parts. Finally, some fully-connected layers are added to map the consistency score and other continuous-valued features into an output score, which indicates the goodness of a resume.}
\label{fig:Model_Architecture}
\end{figure*}

\section{Dataset Description and Model Architecture}
\label{sec:Overview}

\subsection{Resume Dataset}
Since there is no public resume dataset, we build a dataset by collecting resumes from a private resume management company. The resumes are searched results using some queries by experts. Each resume consists of multiple entries, such as user basic information, education level, skills and working experiences. An example is shown in Fig.~\ref{fig:Resume_Example}. For a good resume, an expert may contact the talent, and hence a positive label is assigned. For a relatively bad resume, it may be removed from an expert's searching list and thus we regard it as a negative sample. However, since there are large amounts of resumes returned by the search engine, most of the resumes are not reviewed or the experts do not take any operations on them. Therefore, most of the samples are unlabeled. In particular, we collect $10,343$ resumes in total, while only $33$ and $89$ of them are labeled as positive and negative respectively, and there are $10,221$ unlabeled samples.

\subsection{Resume Quality Assessment Framework}
\begin{figure}[htbp]
\centering
\includegraphics[width=1.0\columnwidth]{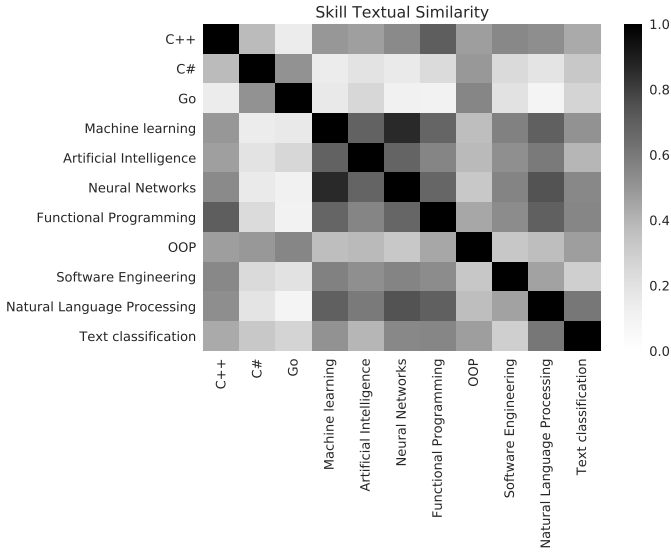}
\caption{Skill similarity scores using the embeddings obtained by the universal sentence encoder \cite{D-Cer-et-al-arXiv-2018}. The semantically similar words (such as ``Machine Learning'' and ``Artificial Intelligence'') tends to have high similarity score.}
\label{fig:Skill_Sim}
\end{figure}

Given the limited labeled and large amounts of unlabeled resumes, our ultimate goal is to learn a model to predict the quality of any new resume. In this paper, we develop a general neural network framework for resume quality prediction, as shown in Fig.~\ref{fig:Model_Architecture}. In particular, we first build a sub-network that measures consistency of a resume's different parts. For example, if a talent claims that he/she has several skills, there should be something (such as working experiences) to support his/her abilities. Hence, the skill part should match well with the working experience part. This can be observed intuitively in Fig.~\ref{fig:Resume_Example}, where the talent claims a ``Micro-services'' skill and then a working experience of ``development of an in-house microservices infrastructure'' is presented. Besides, to apply for a certain job position, different parts (such as skills) of the resume should be in accordance with the requests in the job-post. However, since no job-posts are provided in the dataset, we only estimate the consistency between different parts of the resume. The input of the sub-network can be a set of words or sentences. Either a word or sentence is mapped as a continuous-valued vector using some text embedding techniques. In this paper, we choose the universal sentence encoder \cite{D-Cer-et-al-arXiv-2018} since semantic embeddings can be obtained for either words or sentences, especially for some terminologies. For example, we show the similarities of some skill words embedded by the encoder in Fig.~\ref{fig:Skill_Sim}. From the results, we can see that the different programming language names, ``C++'', ``C\#'' and ``Go'' are close to each other, and also ``Machine Learning'', ``Artificial Intelligence'' and ``Neural Networks'' are highly correlated.

\begin{figure*}[htbp]
\centering
\subfigure[]{\includegraphics[width=0.66\columnwidth]{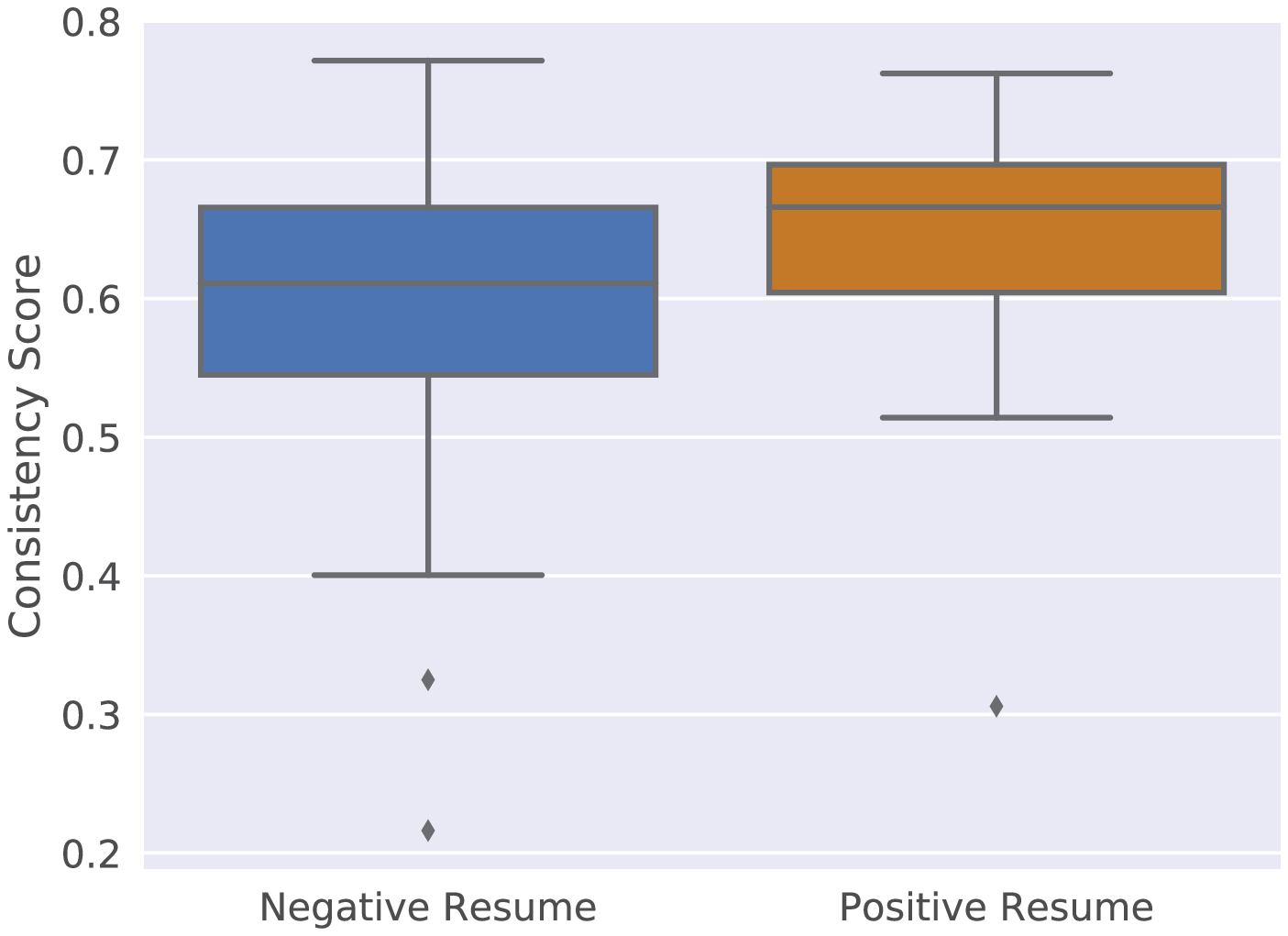}
}
\hfil
\subfigure[]{\includegraphics[width=0.66\columnwidth]{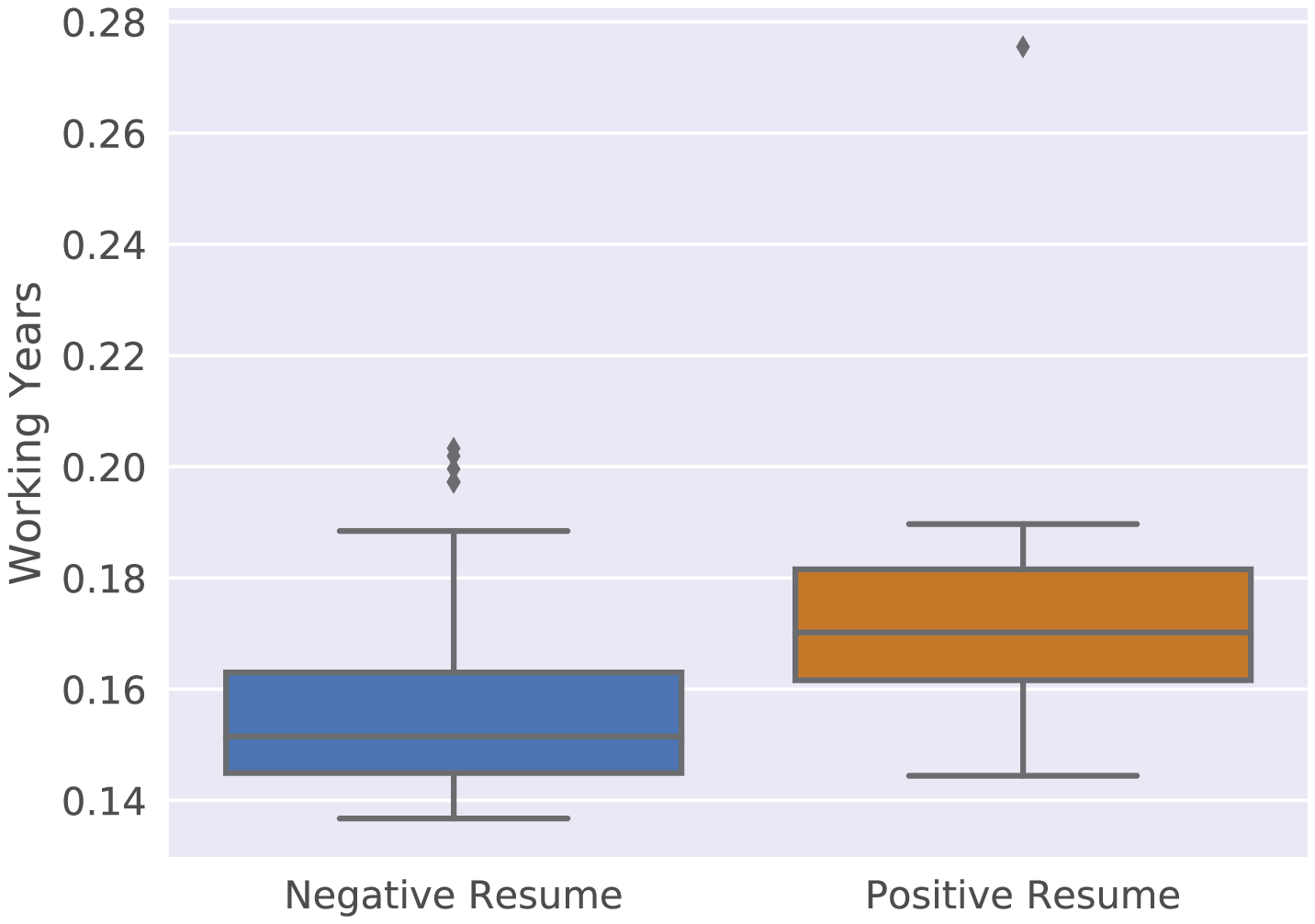}
}
\hfil
\subfigure[]{\includegraphics[width=0.66\columnwidth]{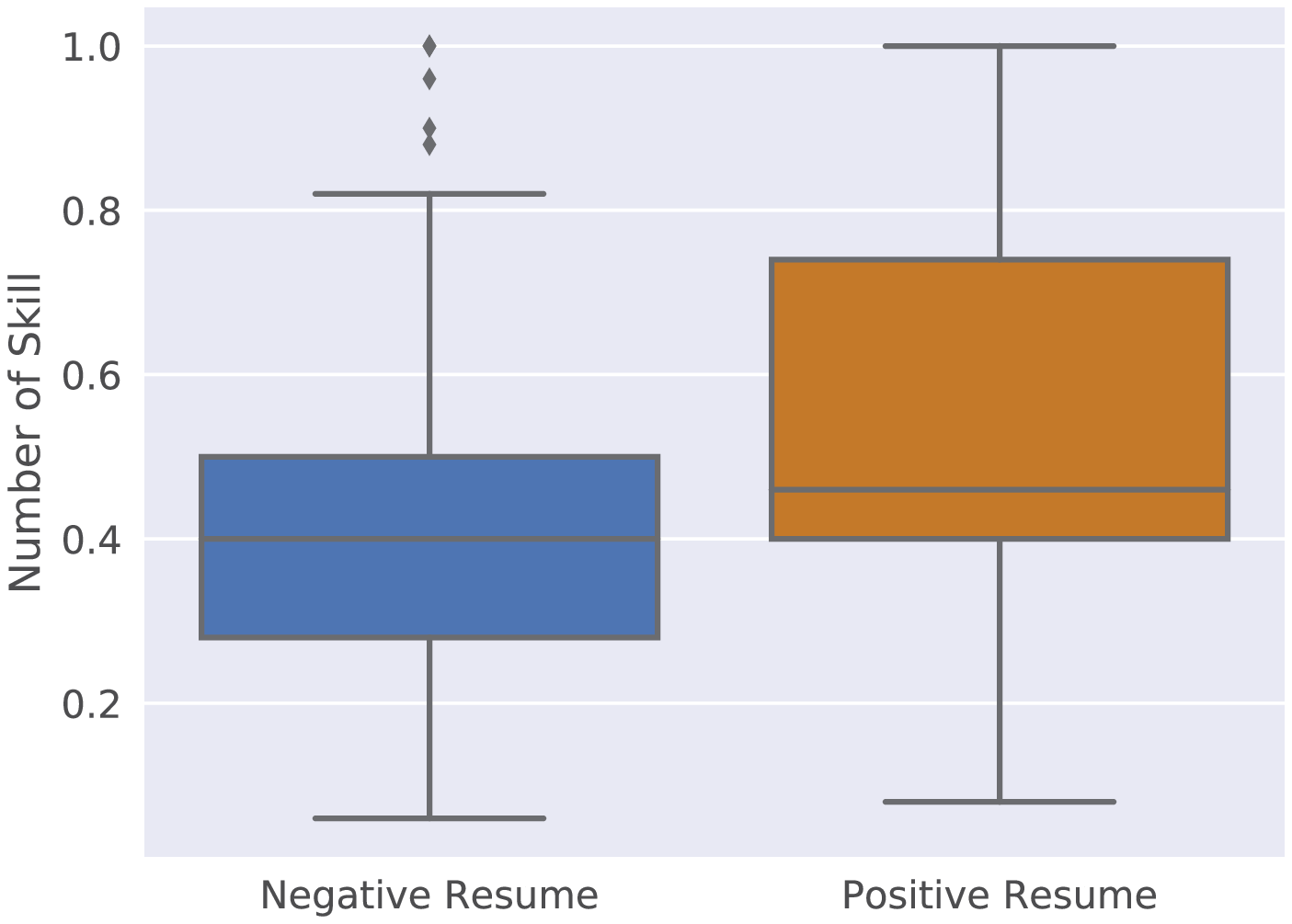}
}
\caption{Statistics of the feature values for the positive and negative resumes: (a) consistency; (b) number of working years; (c) number of skills.}
\label{fig:Fea_Stat}
\end{figure*}

Given the embeddings $\{\mathbf{e}_{mk}\}_{k=1}^{c_m}$ of multiple words/sentences for the $m$-th part of the resume, we learn some adaptive weights $\{\alpha_{mk}\}_{k=1}^{c_m}$ to aggregate them by utilizing the attention scheme \cite{JL-Yang-et-al-CVPR-2017}. Here, $c_m$ is the number of words/sentences. In particular, the weights learning is cast as learning a kernel $\mathbf{q}$, which has the same dimension as the embedding. Then the weight is calculated as:
\begin{equation}
\notag
\beta_{mk} = \mathbf{q}^T \mathbf{e}_{mk},
\end{equation}
and $\alpha_{mk}$ is the normalization of $\beta_{mk}$ using a softmax operator:
\begin{equation}
\notag
\alpha_{mk} = \frac{\exp(\beta_{mk})}{\sum_j \exp(\beta_{mj})}.
\end{equation}
The combined embedding is given by
\begin{equation}
\notag
\mathbf{e}_m = \sum_k \alpha_{mk} \mathbf{e}_{mk}.
\end{equation}
By cascading two attention blocks, larger weights are learned for more discriminative words/sentences. The scheme is particular suitable for our RQA task since it can process arbitrary number of embeddings and is invariant to the embedding order \cite{JL-Yang-et-al-CVPR-2017}. Each combined embedding $\mathbf{e}_m$ is then transformed as a hidden representation $\mathbf{z}_{mH}$ using some fully-connected layers, where $H$ is the number of hidden layers. After that, we calculate the cosine similarity of the different hidden representations $\mathbf{z}_{1H}, \cdots, \mathbf{z}_{MH}$, i.e.,
\begin{equation}
\notag
s(\mathbf{z}_{1H}, \cdots, \mathbf{z}_{MH}) = \frac{M(M-1)}{2} \sum_{m=1,m'>m}^M \frac{\mathbf{z}_{mH}^T \mathbf{z}_{m'H}}{\|\mathbf{z}_{mH}\| \|\mathbf{z}_{m'H}\|}.
\end{equation}
This measures the consistency of different parts. In this paper, we only estimate the consistency between the user skills and working experiences, and thus $M=2$.

\subsection{Other Features}
In addition to the consistency score, we also extract some features that may affect the quality of a resume. These features include:
\begin{itemize}
  \item the number of entries filled in the resume template;
  \item the education level (high school, bachelor, master and doctor);
  \item if the education level is bachelor or higher, the rank of the graduated university;
  \item the number of working years;
  \item the number of awards obtained;
  \item the number of skills; and
  \item the number of previous work positions.
\end{itemize}
In Fig.~\ref{fig:Fea_Stat}, we show statistics of the feature values for positive and negative samples respectively. It can be seen from the results that the feature values of positive and negative samples fall into different ranges. This demonstrates that the extracted features are useful to discriminate good resumes from bad ones.

Finally, the consistency score, as well as the different features are mapped into an output score using some fully-connected layers. A higher score indicates that the resume is more likely to be selected a good candidate for further consideration, such as giving an interview.

\section{Model Formulation and Optimization}
\label{sec:Formulation}
In the section, we first present a baseline model and then some variants to deal with the label deficiency issue.

\subsection{Baseline}
The general formulation of the objective function is given by:
\begin{equation}
\label{eq:General_Objective}
\epsilon(\Theta) = L(\Theta) + \gamma R(\Theta),
\end{equation}
where $L(\Theta)$ is the loss term, $R(\Theta)$ is a regularization term that helps control the model complexity and $\Theta$ is the set of all the parameters to be learned in the model, including the attention kernel $\mathbf{q}$ and the parameters of the fully-connected layers. In the following, we show how to choose the loss and regularization terms.

Given an input sample (resume) $r_i$ and its label $y_i \in \{-1,1\}$ ($1$ indicates good and $-1$ indicates bad), an intuitive idea is to adopt the least squared loss, i.e.,
\begin{equation}
\label{eq:L2_Loss}
L\left(\Theta; (r_i, y_i)\right) = \frac{1}{2} (f_i - y_i)^2,
\end{equation}
where $f_i = f(\Theta; r_i)$ is the output of $r_i$ and a $tanh$ function, i.e.,
\begin{equation}
\notag
tanh(x) = \frac{e^x - e^{-x}}{e^x + e^{-x}}
\end{equation}
is added to map the output into $[-1,1]$. In this baseline model, we do not learn the attention kernel to reduce the number of parameters to be learned since there is only a few labeled samples. The regularization term $R(\Theta)$ consists of some Frobenius-norm based regularization terms, e.g., $\frac{\gamma}{2} \| W \|_F^2$ to reduce the chance of over-fitting. Here, $W$ is the weight parameter of a fully-connected layer and $\gamma \geq 0$ is a trade-off hyper-parameter.

\subsection{Handling Limited Labeled Data}

\subsubsection{Varying the Loss}
To alleviate the label deficiency issue and also deal with the class-imbalance problem, we propose two variants of the model by changing the loss:
\begin{itemize}
  \item In the first variant, we utilize the contrastive loss \cite{S-Chopra-et-al-CVPR-2005}, where the input is a pair of samples $(r_i^1, r_i^2)$. The label of the pair is $y_i=1$ or $0$, which indicates whether the two samples are from the same category or not. The loss function is given by
      \begin{equation}
      \label{eq:Contrast_Loss}
      L\left(\Theta; (r_i^1, r_i^2, y_i)\right) = y_i \frac{2}{\eta} \delta_i^2 + (1 - y_i) 2\eta \mathrm{\exp}(-\frac{2.77}{\eta} \delta_i),
      \end{equation}
      where $\delta_i = f(\Theta; r_i^1) - f(\Theta; r_i^2)$ and we set the hyper-parameter $\eta = 2$, which is the upper bound of $\delta_i$. It should be noted that when $y_i=0$, $r_i^1$ should be the positive sample and $r_i^2$ is negative since we want the outputs of positive samples to be larger than negative ones.
  \item Alternatively, we can adopt the triplet loss \cite{M-Norouzi-et-al-NIPS-2012, F-Schroff-et-al-CVPR-2015}, where the input is a triplet of samples $(r_i^a, r_i^p, r_i^n)$. Here, $r_i^a$ is an anchor positive sample, $r_i^p$ is another positive sample and $r_i^n$ is any negative sample. The loss function for the triplet is:
      \begin{equation}
      \label{eq:Triplet_Loss}
      L\left(\Theta; (r_i^a, r_i^p, r_i^n)\right) = [|f_i^a - f_i^p|  - (f_i^a - f_i^n) + \mu]_+,
      \end{equation}
      Here, $[\rho]_+ \equiv \mathrm{max}(0, \rho)$ and $\mu$ is a hyper-parameter to be determined. It should be noted that we do not calculate the absolute difference of $f_i^a$ and $f_i^n$ since we want penalize the case that the negative samples have large prediction scores.
\end{itemize}

\subsubsection{Leveraging Unlabeled Data}
Alternatively, we can make use of the large amounts of unlabeled data to deal with label deficiency issue. This is conducted by adding a manifold regularization (MR) \cite{M-Belkin-et-al-JMLR-2006} term, which can exploit the structure of the data distribution. The term is given as follows:
\begin{equation}
\label{eq:MR_Term}
R_I (\Theta; \{r_i, r_j^u\}) = \gamma_I \omega_{ij} (f_i - f_j^u)^2,
\end{equation}
where $f_i$ and $f_j^u$ are outputs of a labeled and unlabeled samples $r_i$ and $r_j^u$ respectively, $\omega_{ij}$ is the similarity of the two samples and $\gamma_I$ is a trade-off hyper-parameter. Here, $\omega_{ij}$ is calculated using the following strategy: for the skills and working experiences, we compute their respective average embeddings. Then the cosine similarity of the two embeddings, together with the other features are concatenated as a vector $\mathbf{v}$, and the similarity of the two samples is
\begin{equation}
\label{eq:Similarity}
\begin{split}
\omega_{ij} =
\left\{
\begin{array}{cc}
\mathrm{\exp} \left( -\frac{\|\mathbf{v}_i - \mathbf{v}_j^u\|_2^2}{2 \sigma^2} \right), & r_j^u \in NB_k(r_i); \\
0, & \mathrm{otherwise}.
\end{array}
\right.
\end{split}
\end{equation}
where $NB_k(r_i)$ is the set of $k$-nearest neighbor of sample $r_i$, $\sigma$ is the bandwidth hyper-parameter and we set it as $1$ empirically.

Intuitively, if two samples have similar features, their prediction scores should be close. This enables the model to be smoothed along the data manifold and helps reduce the model complexity \cite{Y-Luo-et-al-TIP-2013}. By adding the regularization term (\ref{eq:MR_Term}) to the baseline model (\ref{eq:L2_Loss}), geometry of the data distribution can be well exploited \cite{M-Belkin-et-al-JMLR-2006} and hence better performance can be achieved, especially when the labeled data are scarce.

\subsection{Optimization}
Based on the designed objective functions, we adopt the stochastic gradient descent (SGD) method for optimization, i.e.,
\begin{equation}
\Theta \leftarrow \Theta - \lambda \frac{\partial \epsilon(\Theta)}{\partial \Theta},
\end{equation}
where $\lambda$ is the learning rate, which is set as $0.01$ empirically in this paper. In each iteration, only one data point (single sample, sample pair or triple) is selected to update the model parameters since the number of input skills and working experiences vary for different resumes. Hence it is hard to process multiple resumes simultaneously at each time.

\subsection{Implementation Details}
After investigating the dataset, we found that a talent usually has only one or a few working experiences. Hence, we directly compute the average embedding of working experiences without attention, and the attention mechanism is only adopted to aggregate multiple skill embeddings.

In this paper, since the same encoder is applied to embed the words and sentences, their embeddings have the same dimensions and statistic properties. Therefore, we do not transform $\mathbf{e}_m$ as hidden representation but directly set $\mathbf{z}_{mH} = \mathbf{e}_m$. If different embedding techniques are adopted, the hidden layers should be added to map the heterogeneous embeddings as comparable representations. All the dimensions of different embeddings and the attention kernel vector $\mathbf{q}$ are $512$. The initialization of the weight parameters in the fully-connected layers are normally distributed random values within the range $[-1, 1]$.

\begin{figure*}[htbp]
\centering
\subfigure[]{\includegraphics[width=0.66\columnwidth]{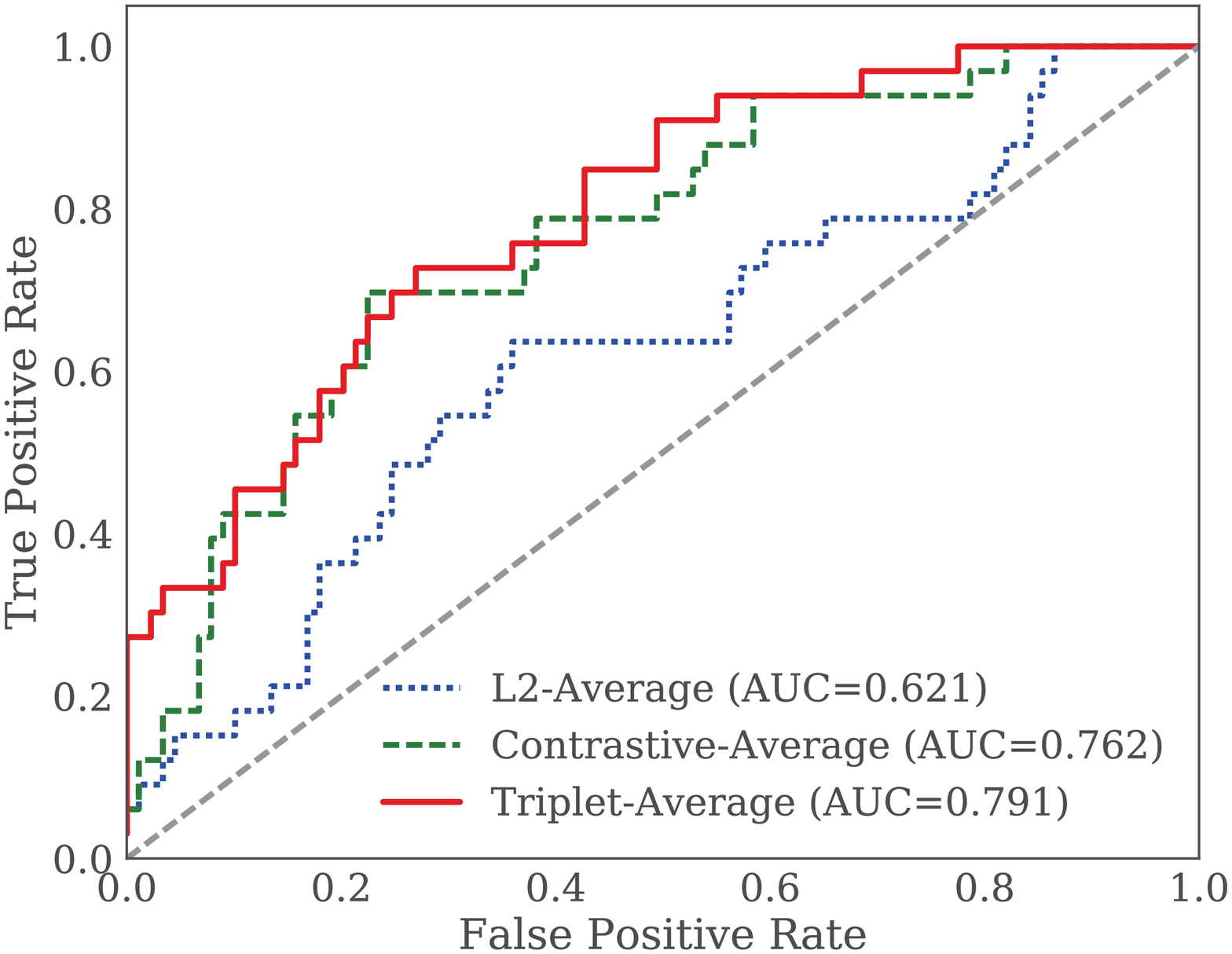}
}
\hfil
\subfigure[]{\includegraphics[width=0.66\columnwidth]{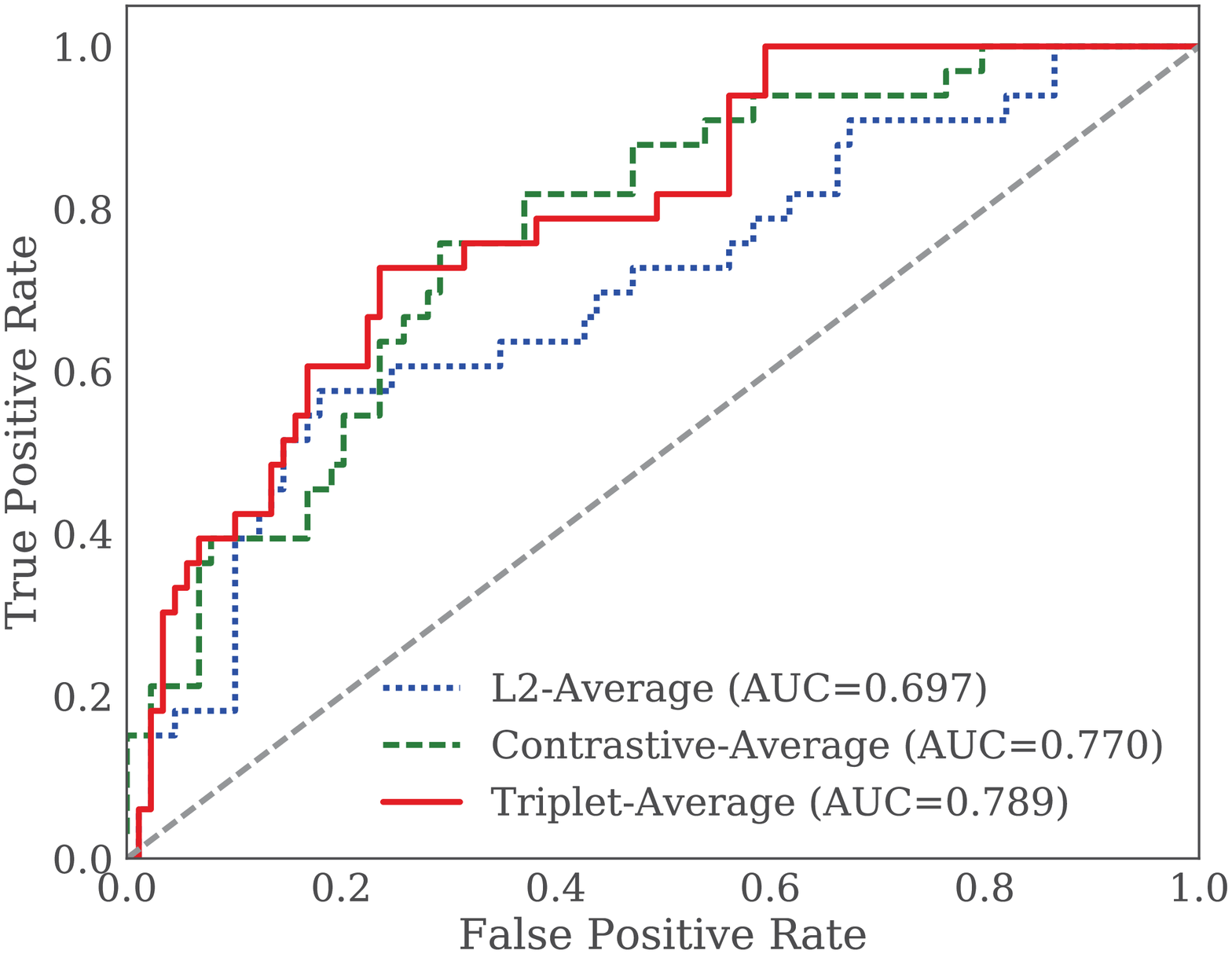}
}
\hfil
\subfigure[]{\includegraphics[width=0.66\columnwidth]{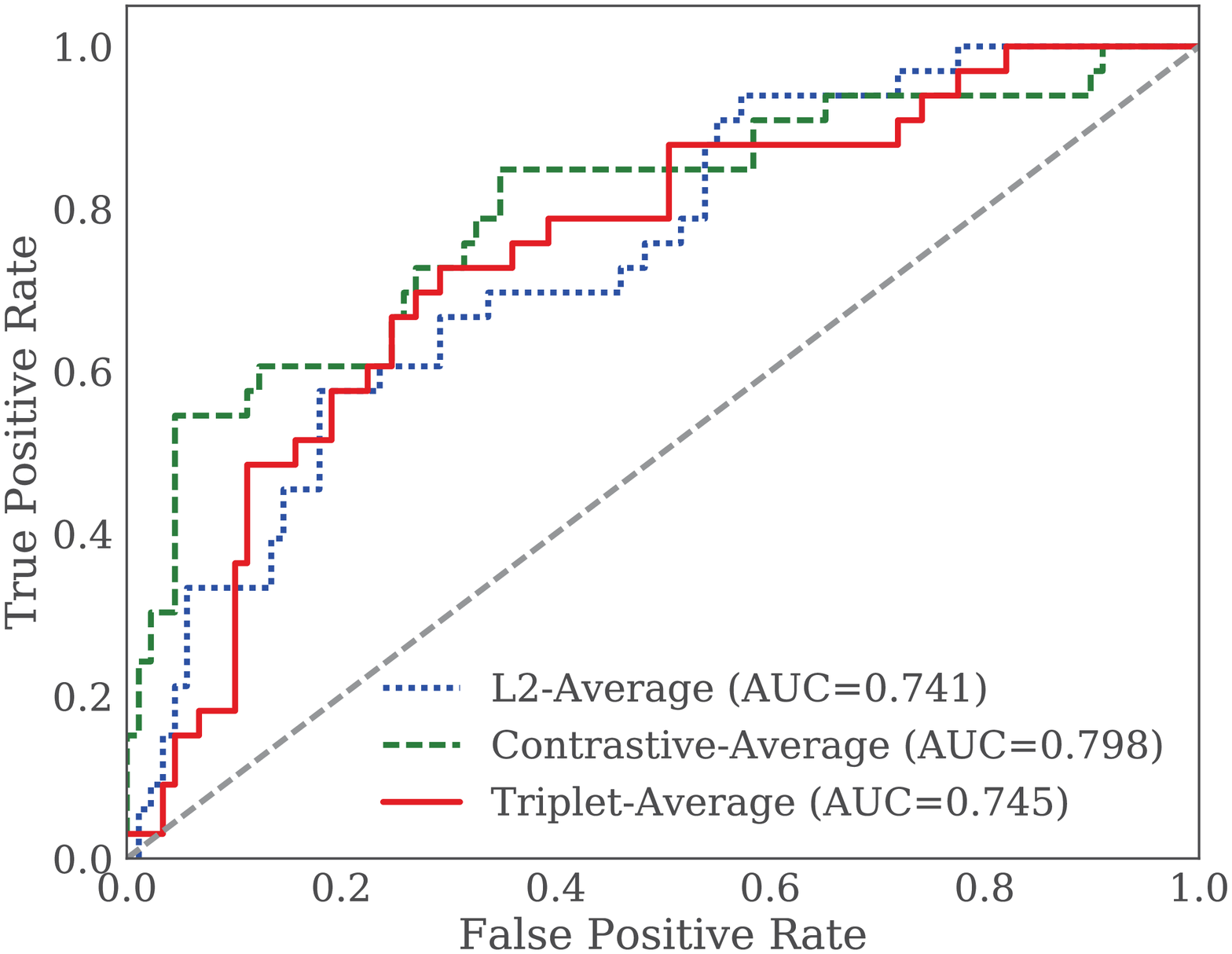}
\label{fig:ROC_AVG_3}
}
\hfil
\subfigure[]{\includegraphics[width=0.66\columnwidth]{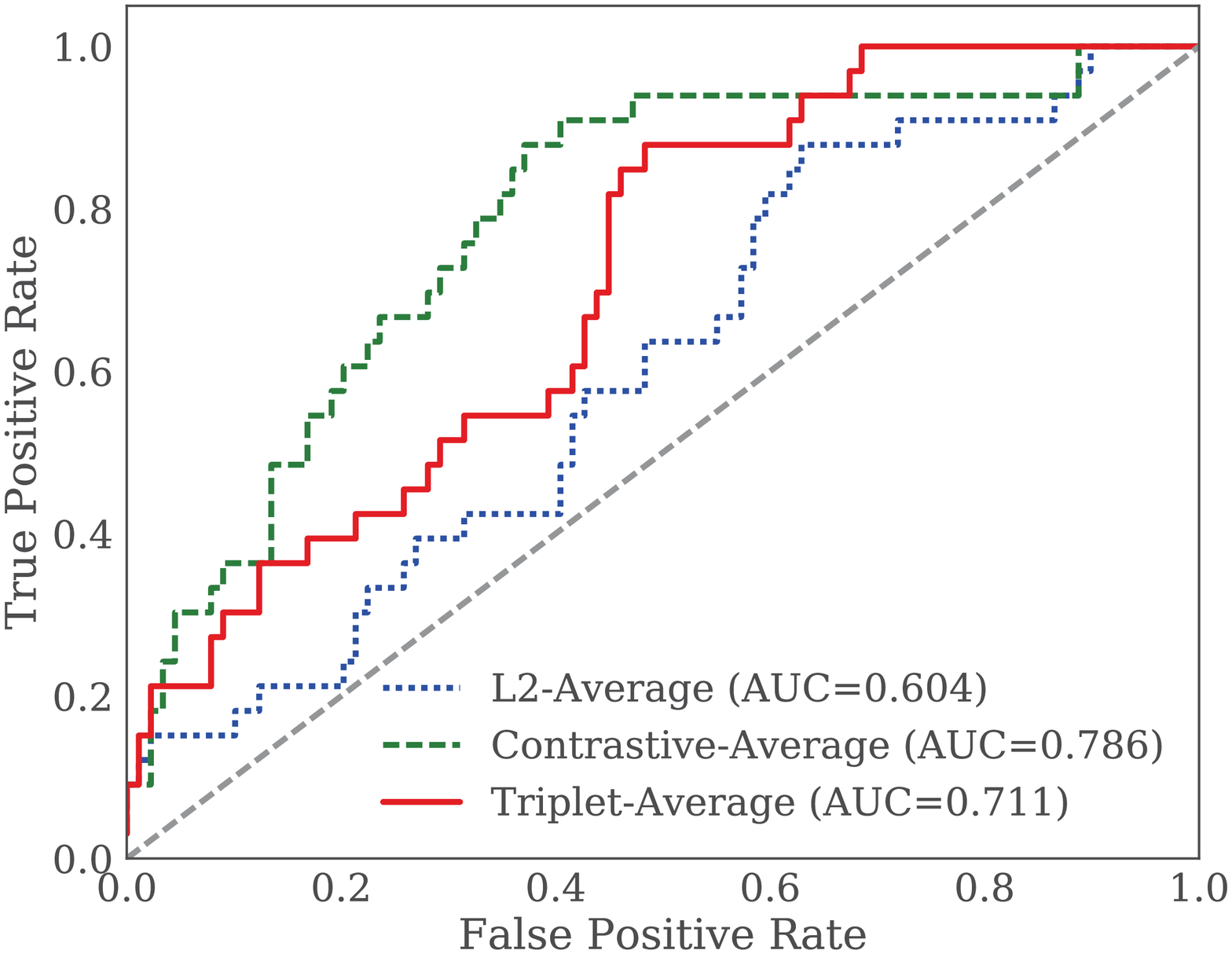}
\label{fig:ROC_AVG_4}
}
\hfil
\subfigure[]{\includegraphics[width=0.66\columnwidth]{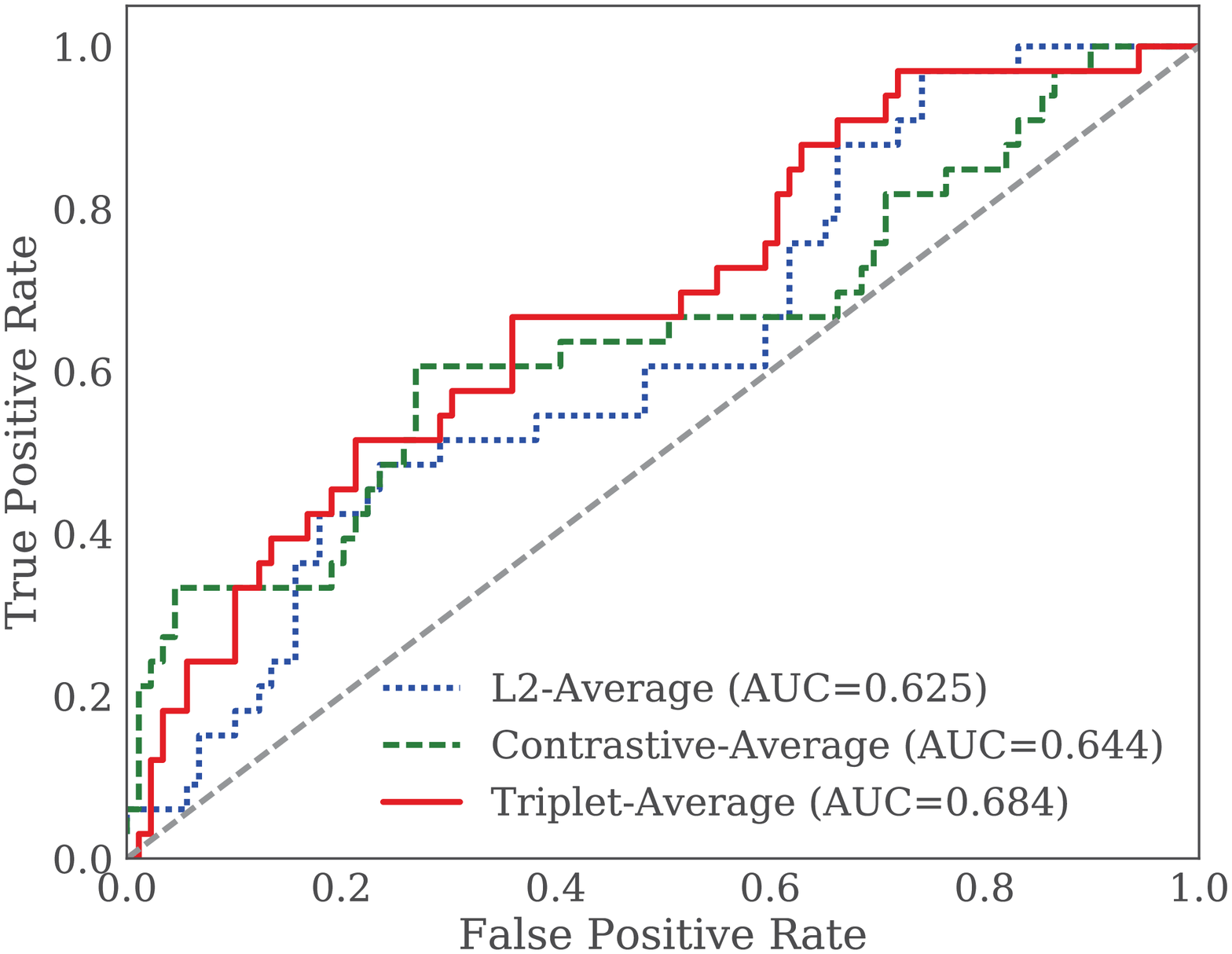}
}
\hfil
\subfigure[]{\includegraphics[width=0.66\columnwidth]{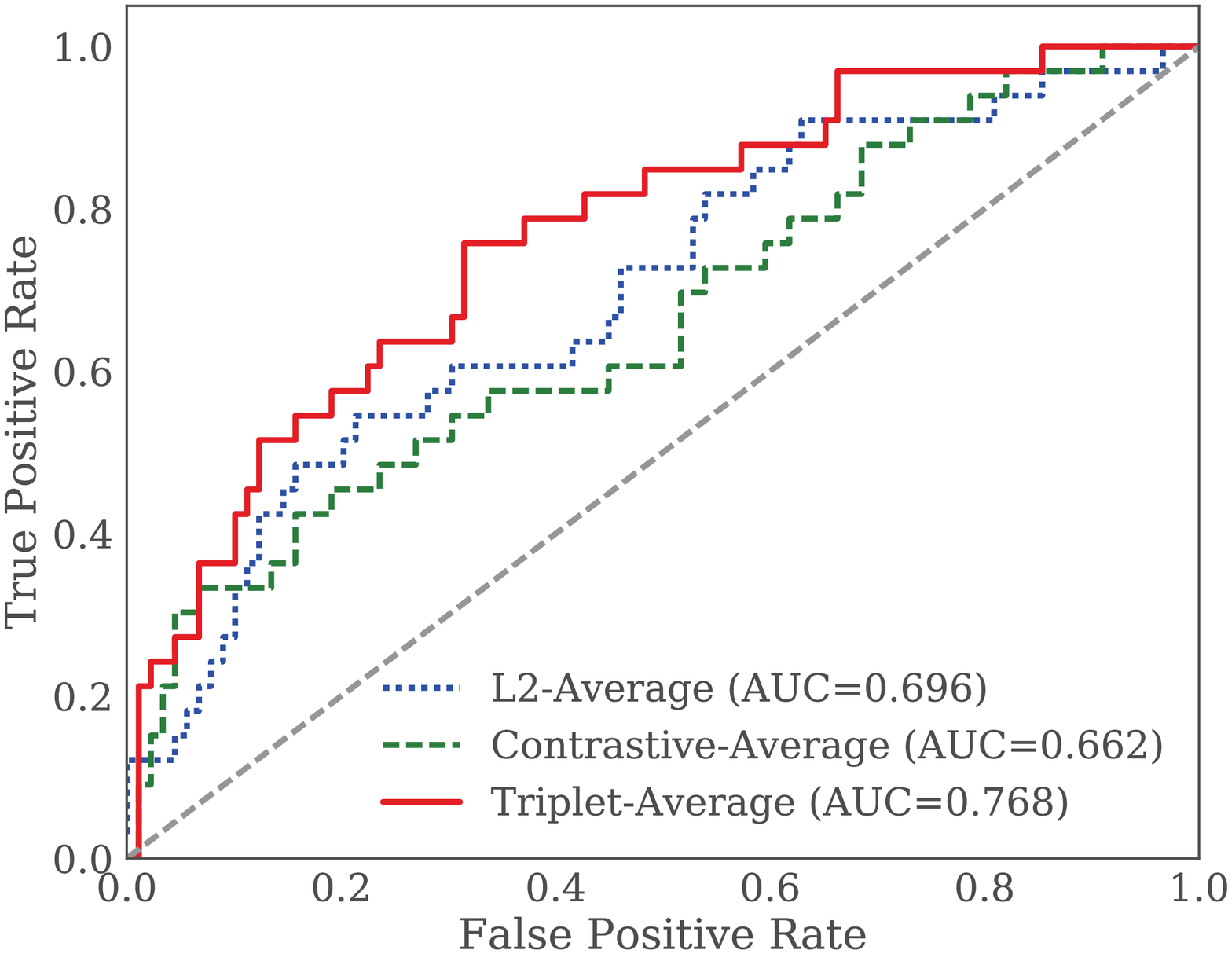}
\label{fig:ROC_AVG_6}
}
\hfil
\subfigure[]{\includegraphics[width=0.48\columnwidth]{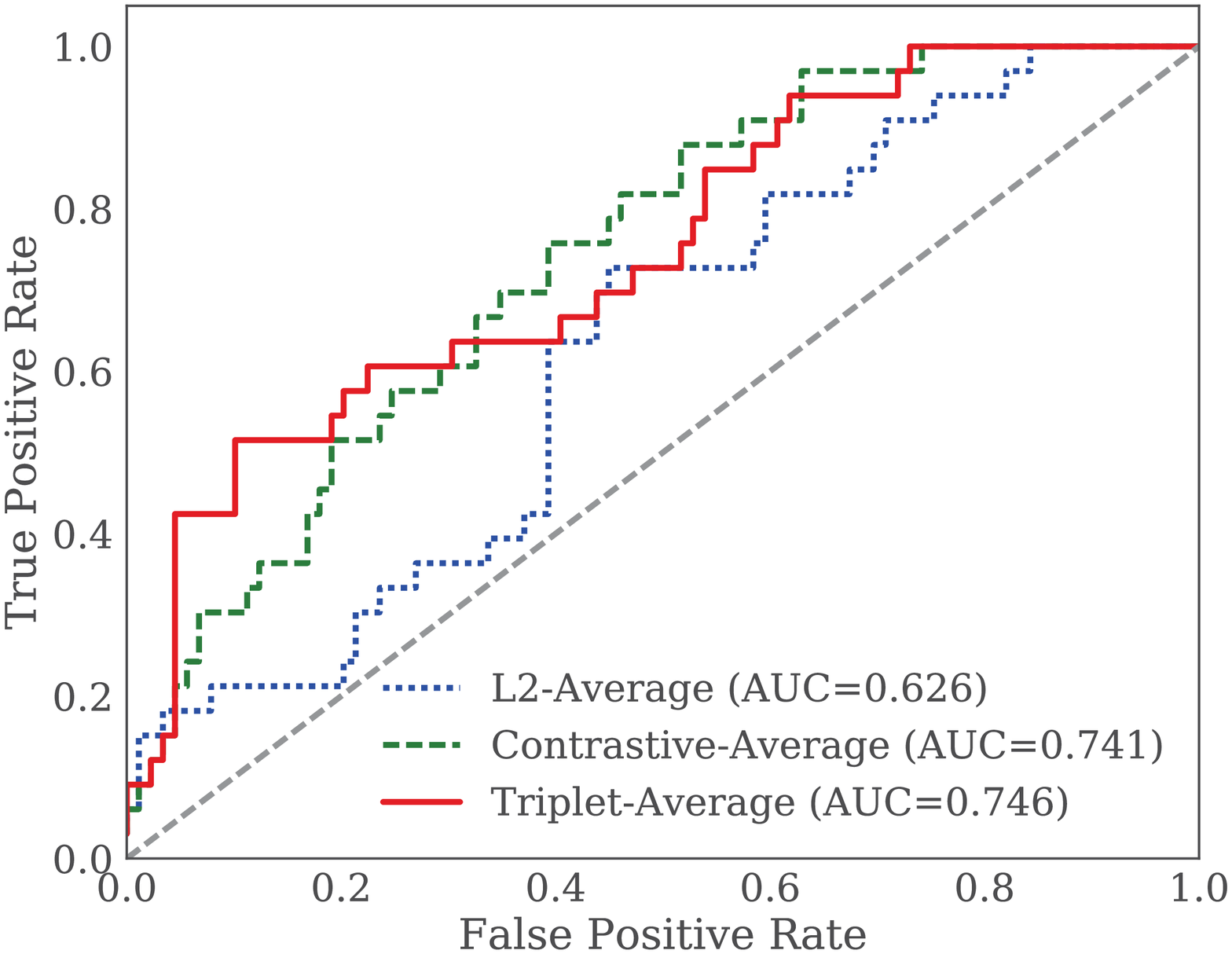}
}
\hfil
\subfigure[]{\includegraphics[width=0.48\columnwidth]{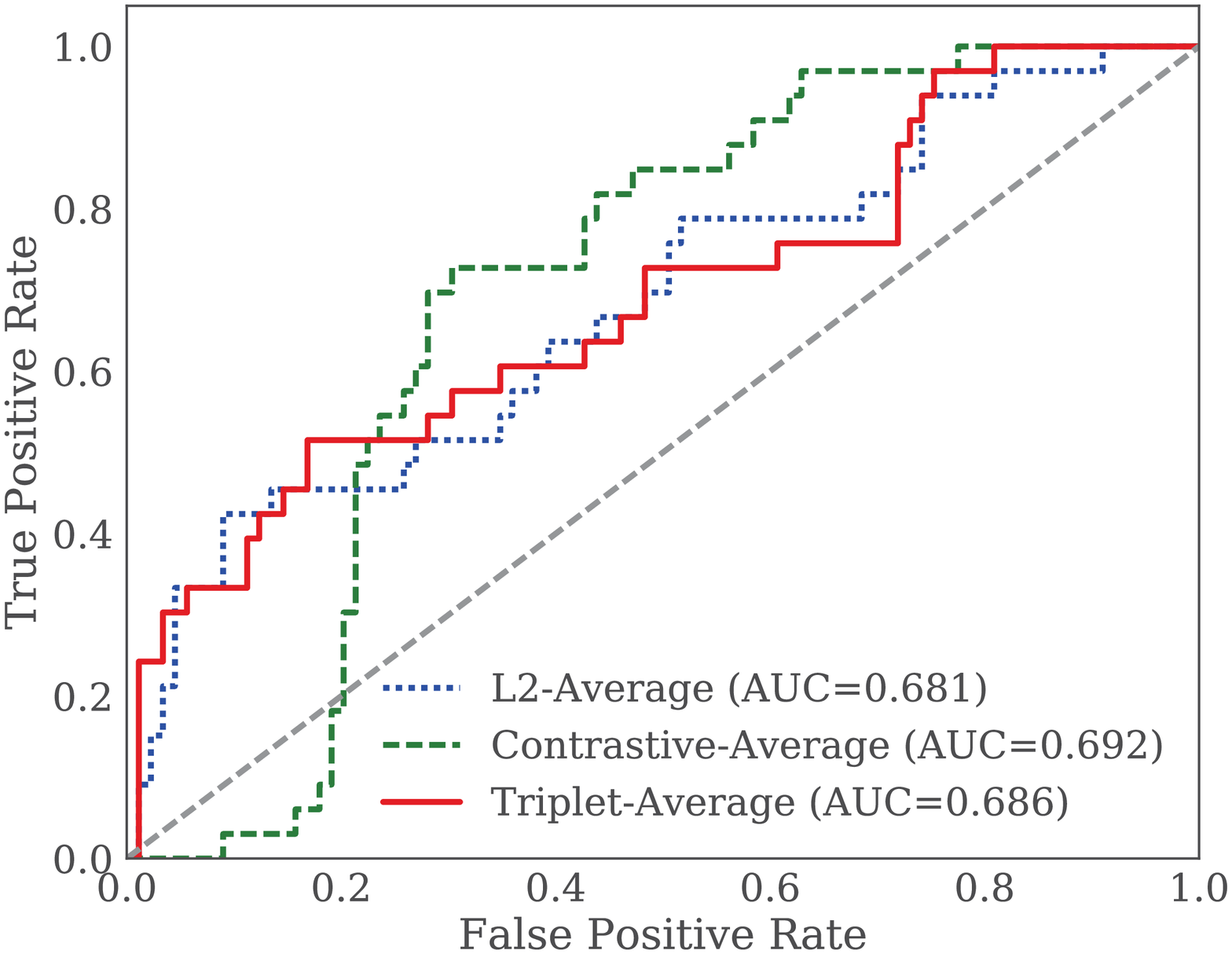}
}
\hfil
\subfigure[]{\includegraphics[width=0.48\columnwidth]{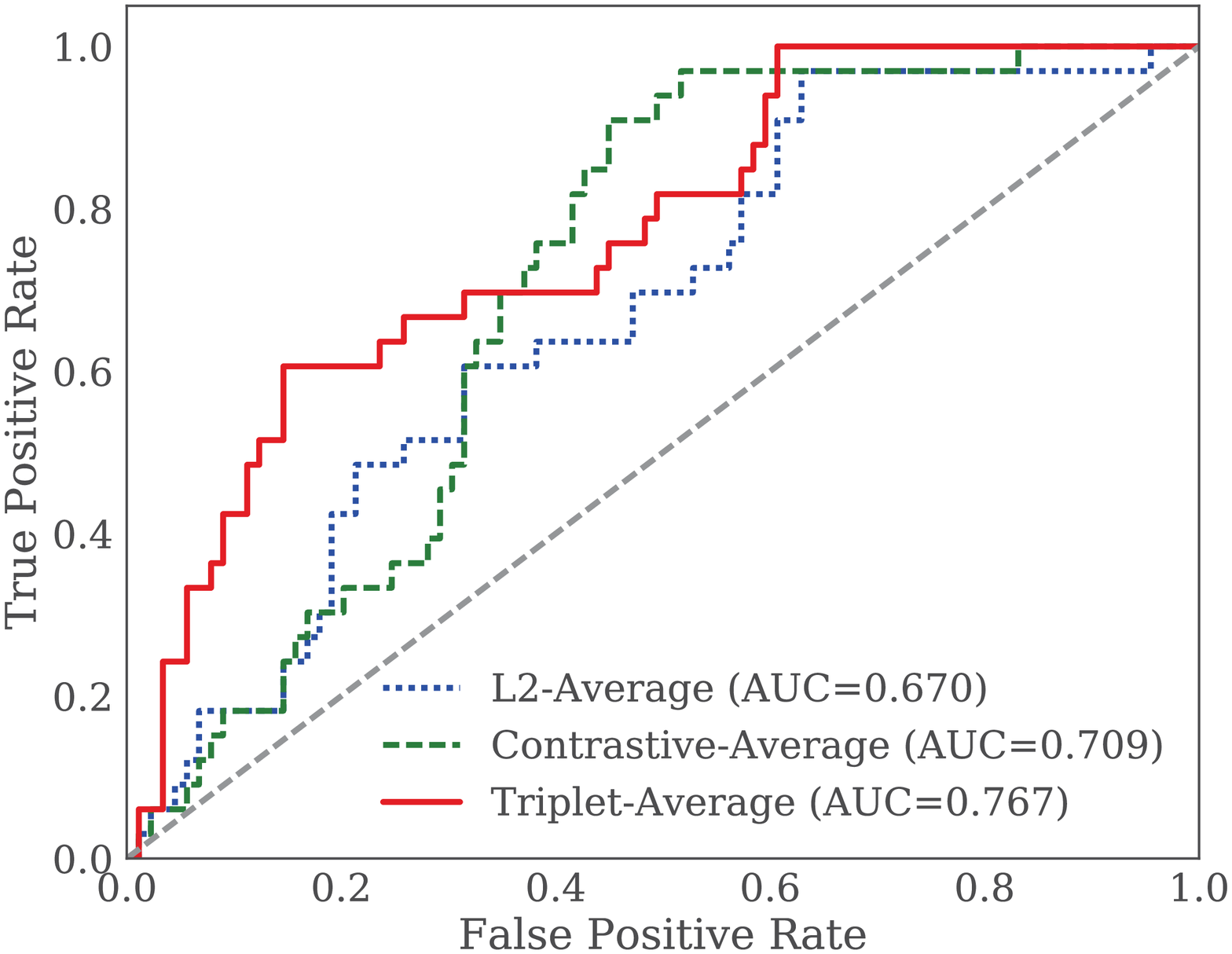}
}
\hfil
\subfigure[]{\includegraphics[width=0.48\columnwidth]{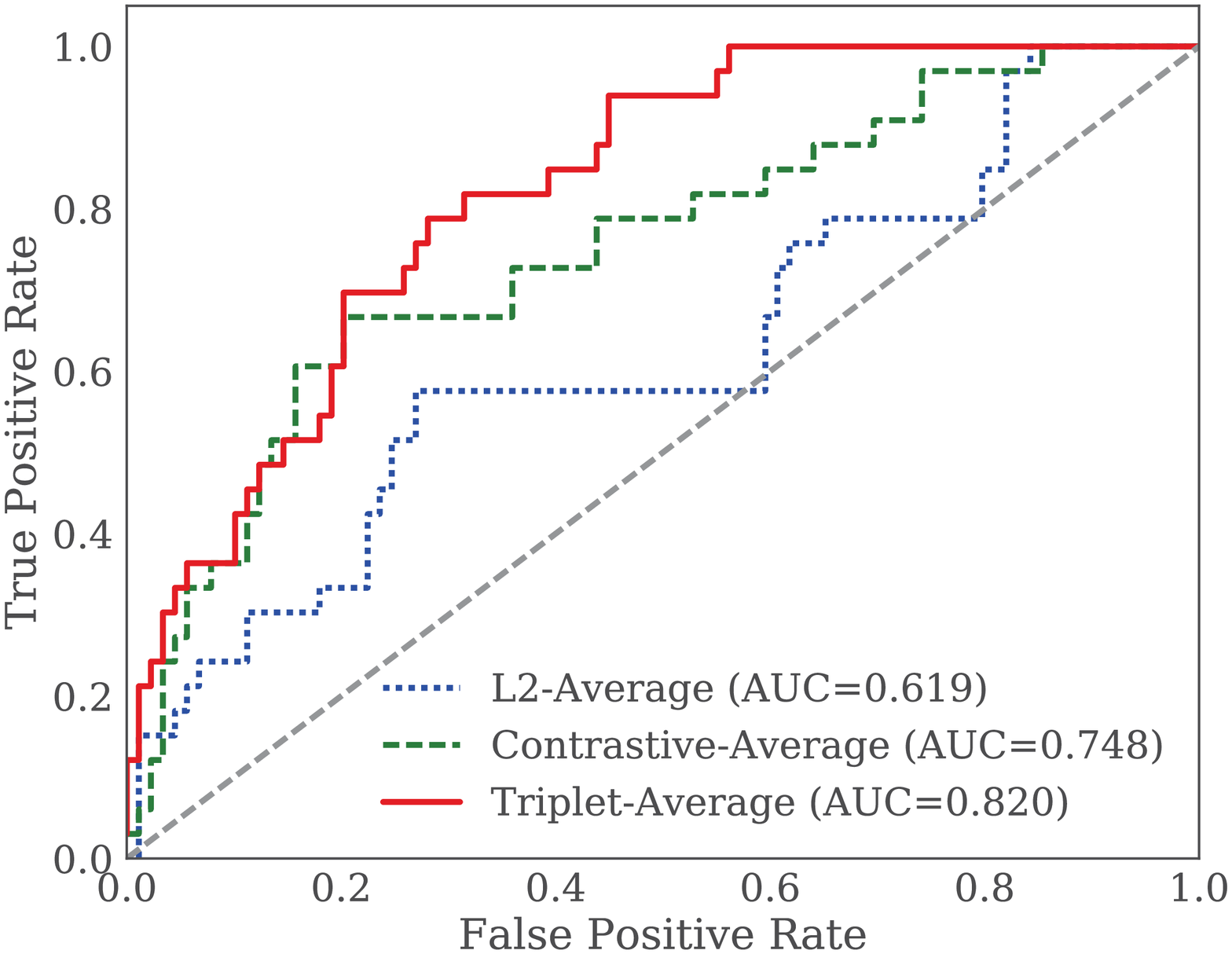}
}
\caption{ROC curves for the $10$ random shuffles of the labeled set (embeddings of multiple skills are averaged merely without using the attention scheme).}
\label{fig:ROC_AVG}
\end{figure*}

\section{Experiments}
\label{sec:Experiments}
In this section, we evaluate the proposed resume quality assessment (RQA) model and its several variants on the established dataset. We first compare the performance of utilizing different losses in the proposed framework, and then investigate the performance of leveraging unlabeled data. Finally, we compare our results with the ones obtained from a resume scoring website.

\begin{figure*}[htbp]
\centering
\subfigure[]{\includegraphics[width=0.66\columnwidth]{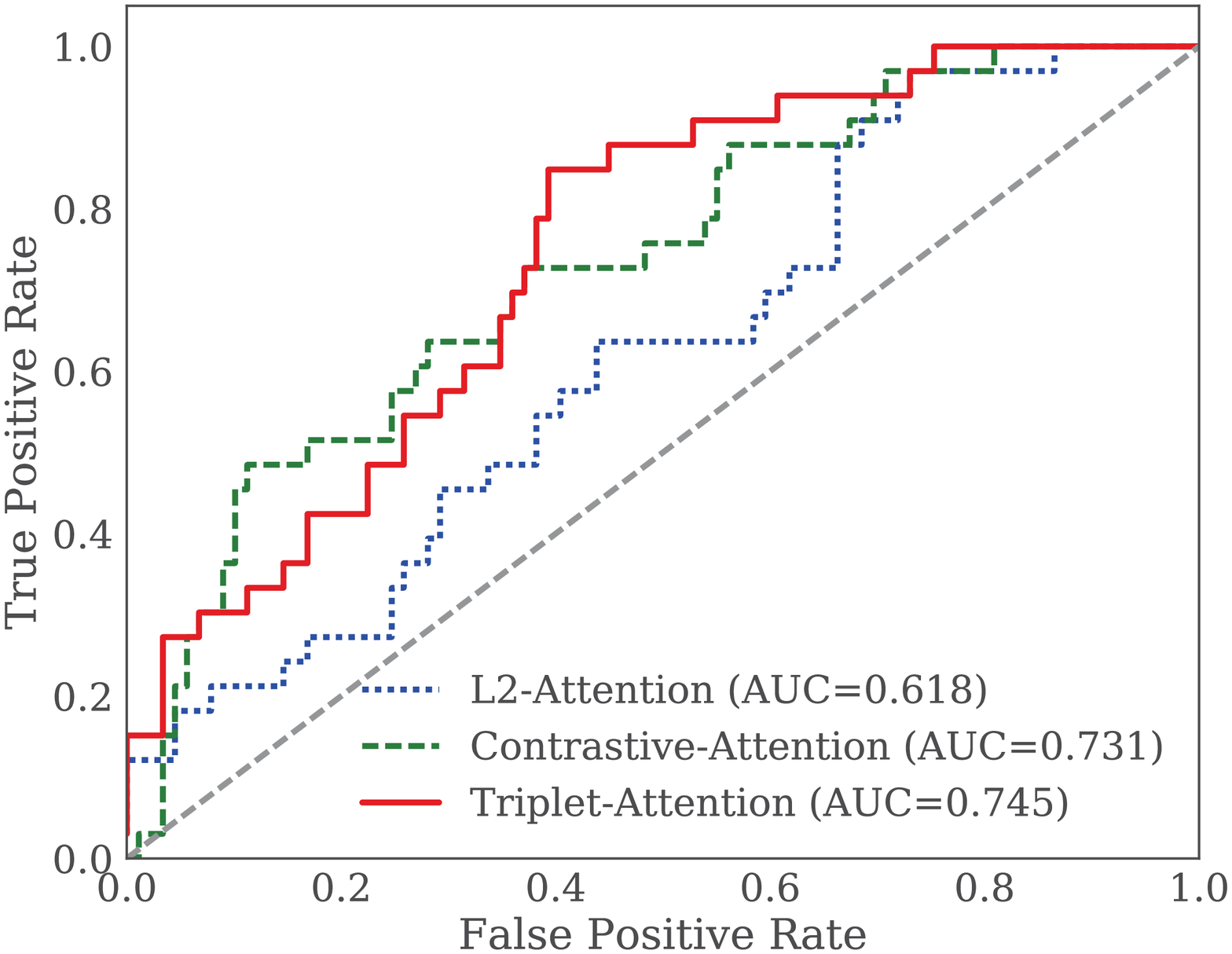}
}
\hfil
\subfigure[]{\includegraphics[width=0.66\columnwidth]{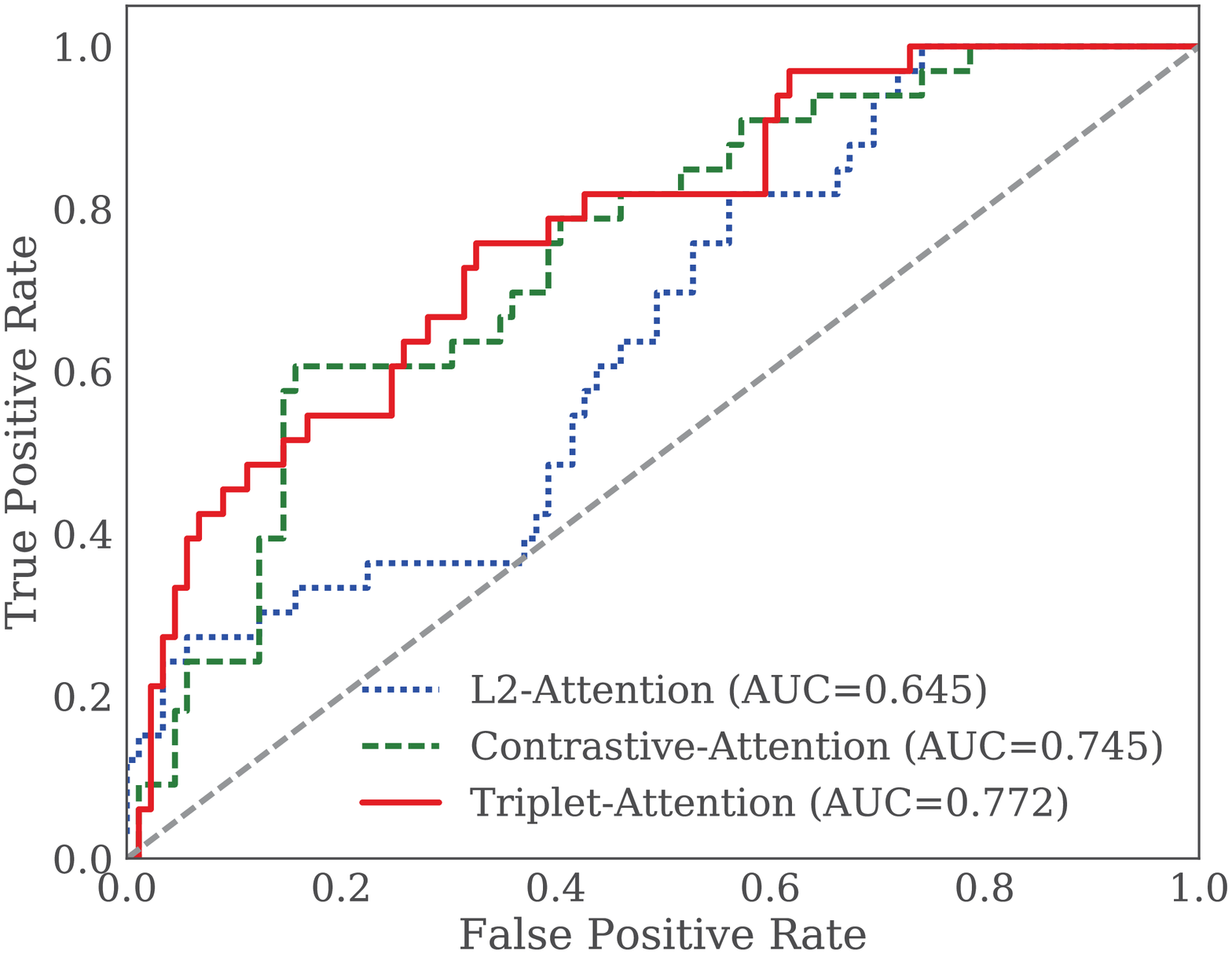}
}
\hfil
\subfigure[]{\includegraphics[width=0.66\columnwidth]{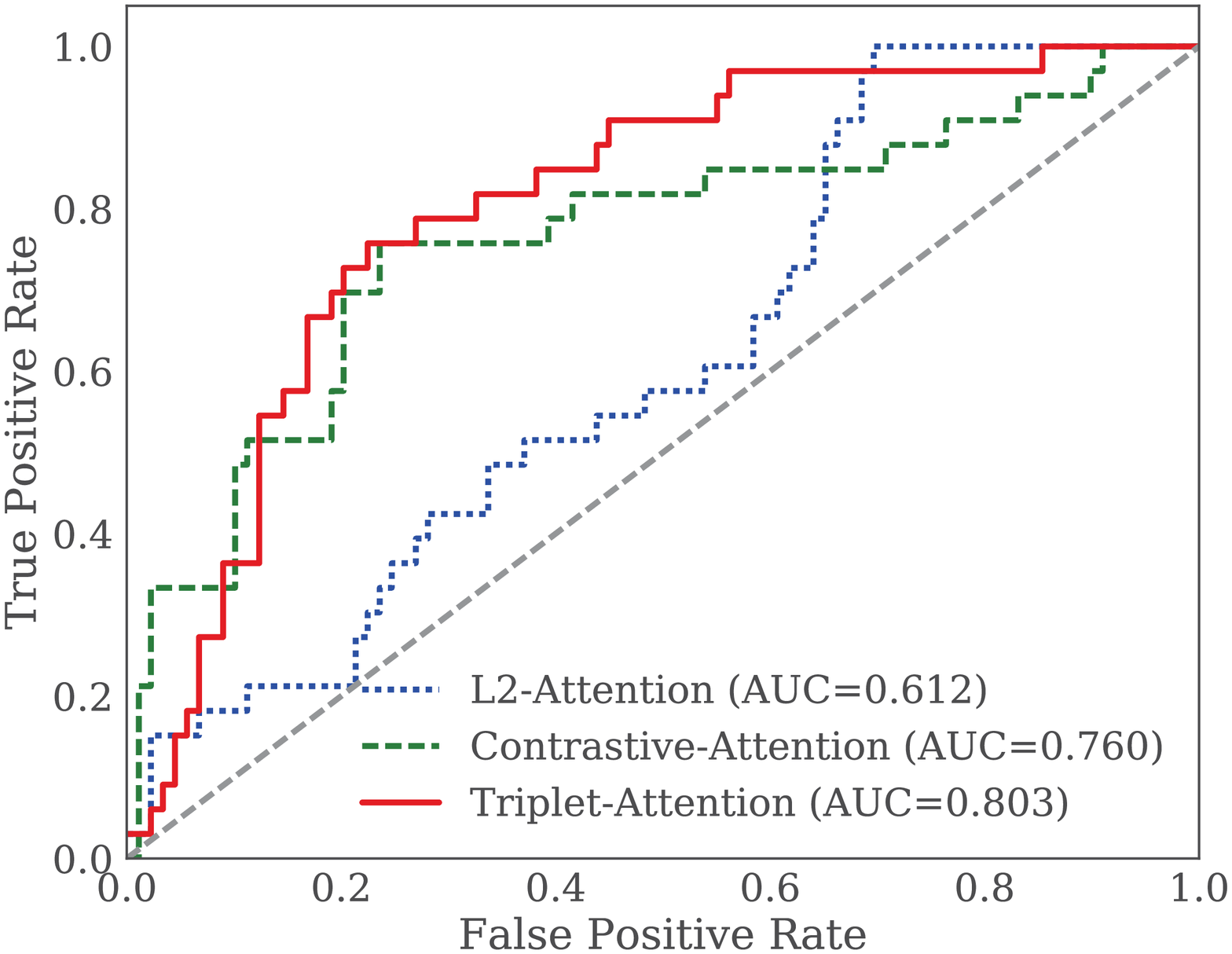}
}
\hfil
\subfigure[]{\includegraphics[width=0.66\columnwidth]{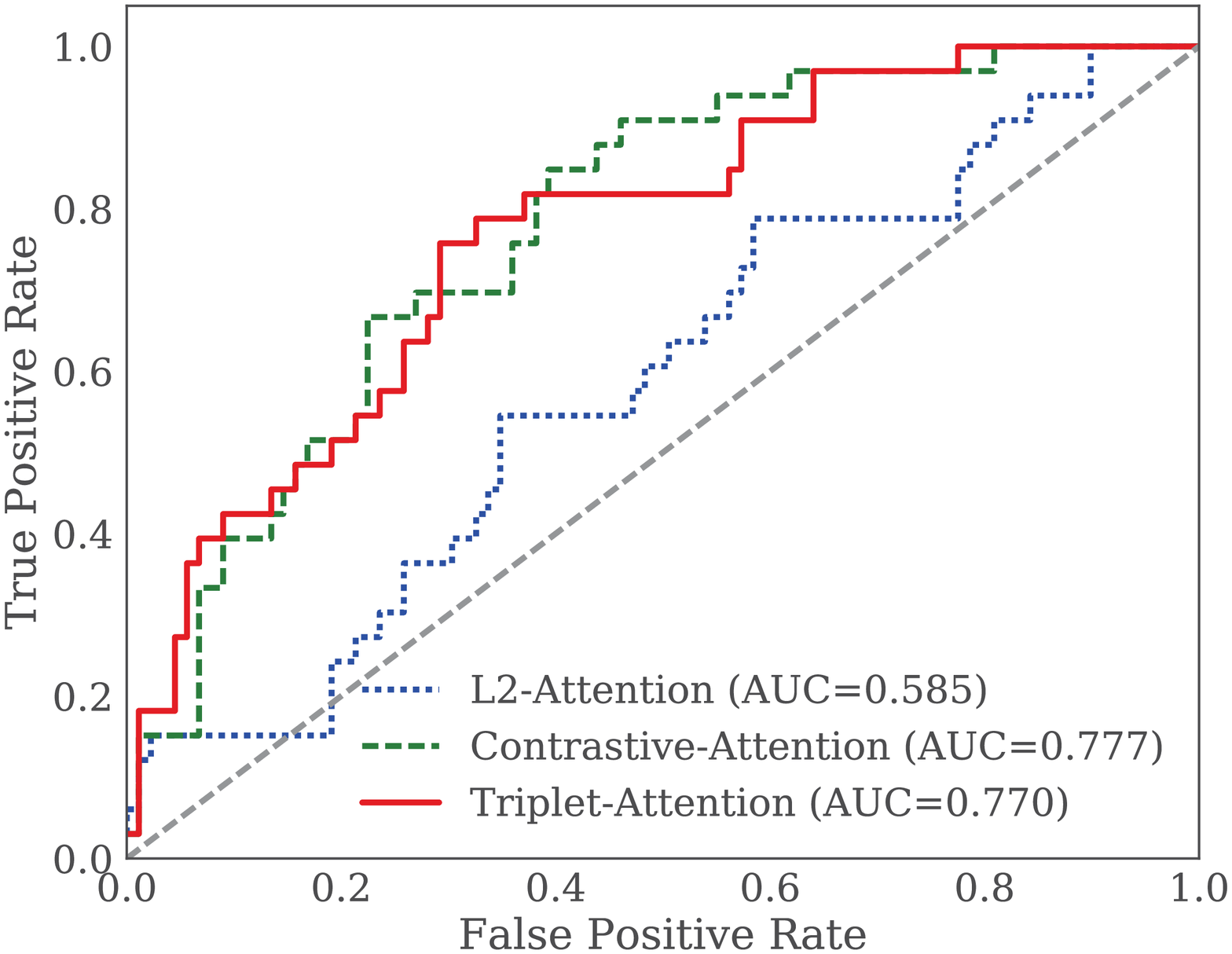}
}
\hfil
\subfigure[]{\includegraphics[width=0.66\columnwidth]{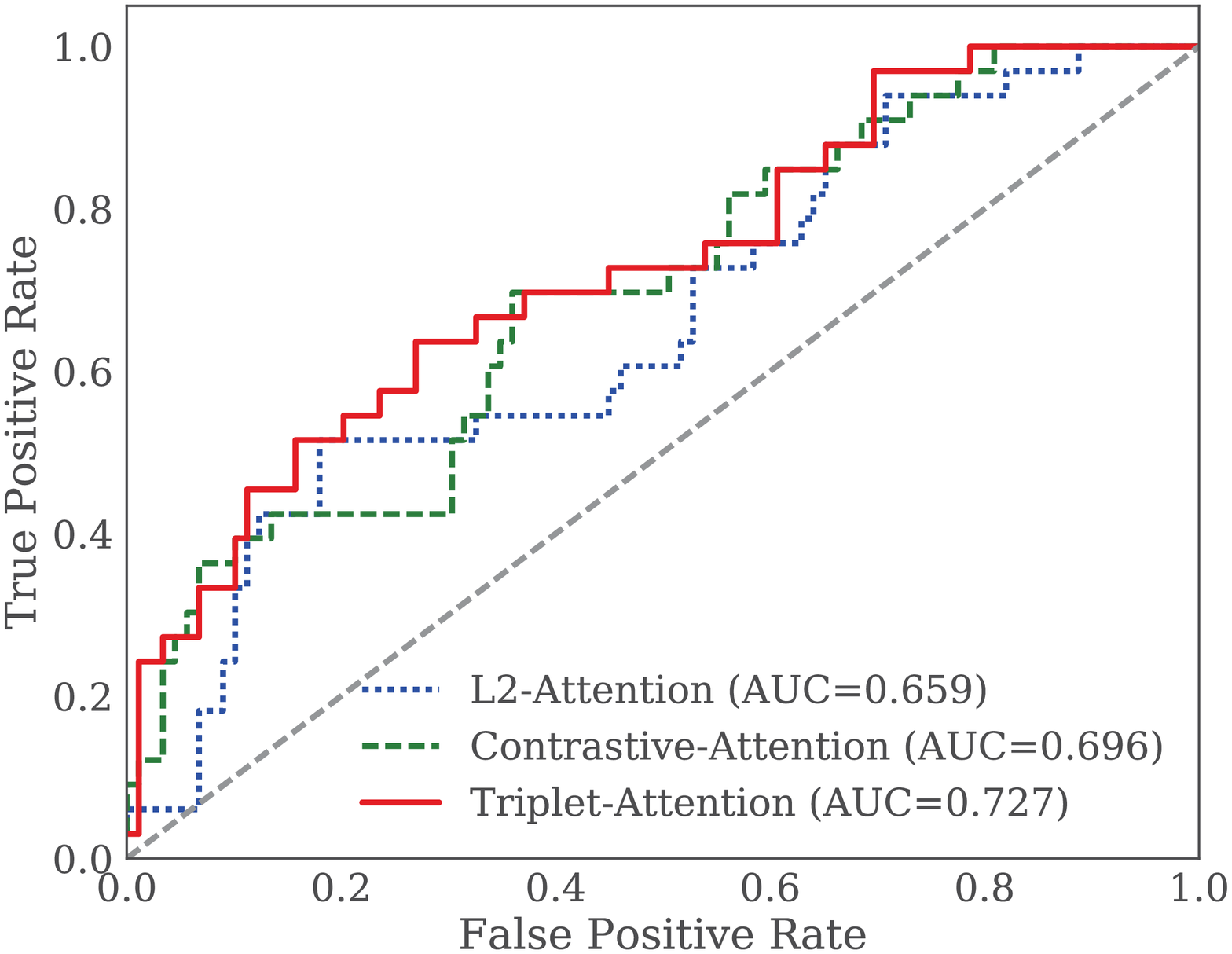}
}
\hfil
\subfigure[]{\includegraphics[width=0.66\columnwidth]{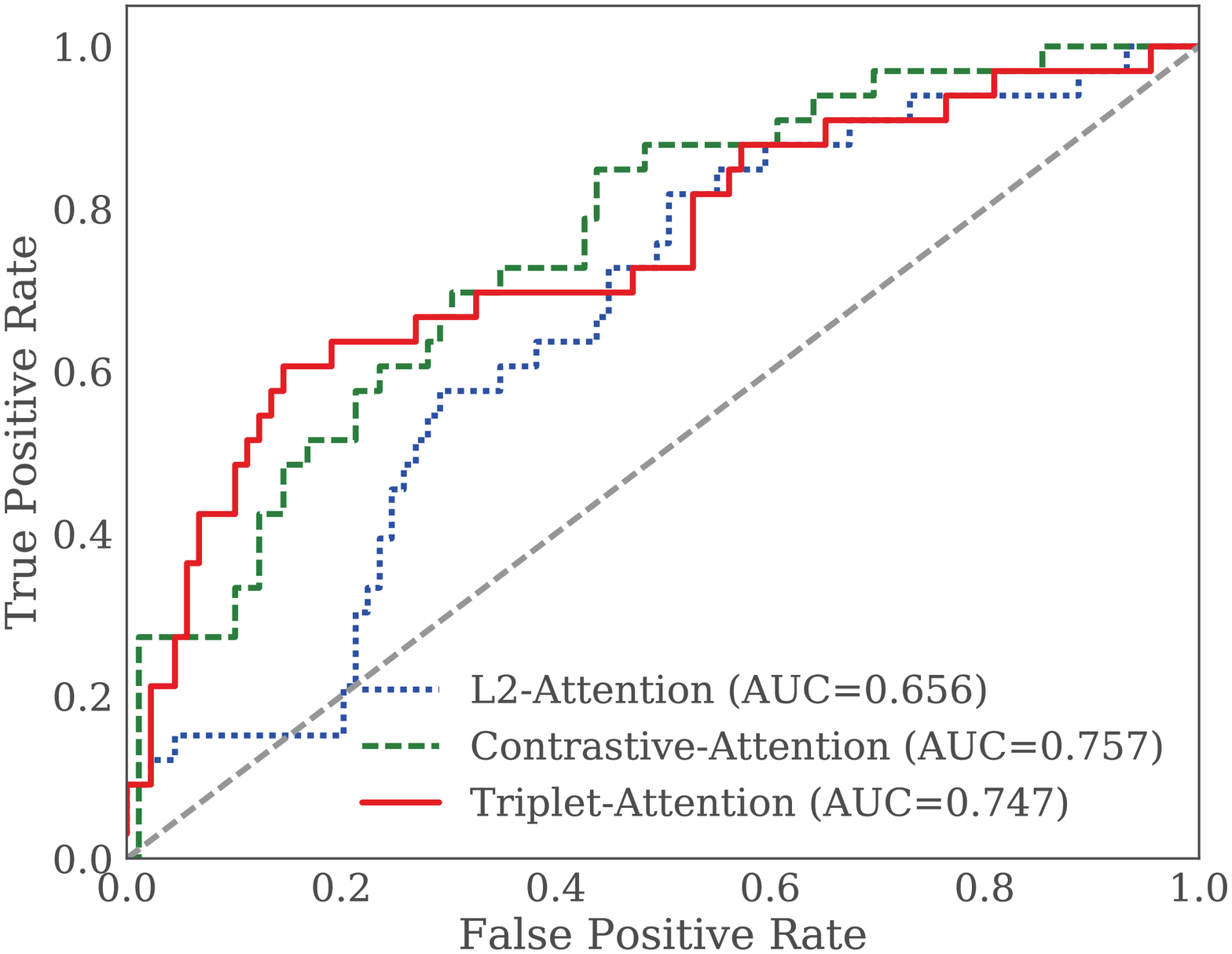}
\label{fig:ROC_ATT_6}
}
\hfil
\subfigure[]{\includegraphics[width=0.48\columnwidth]{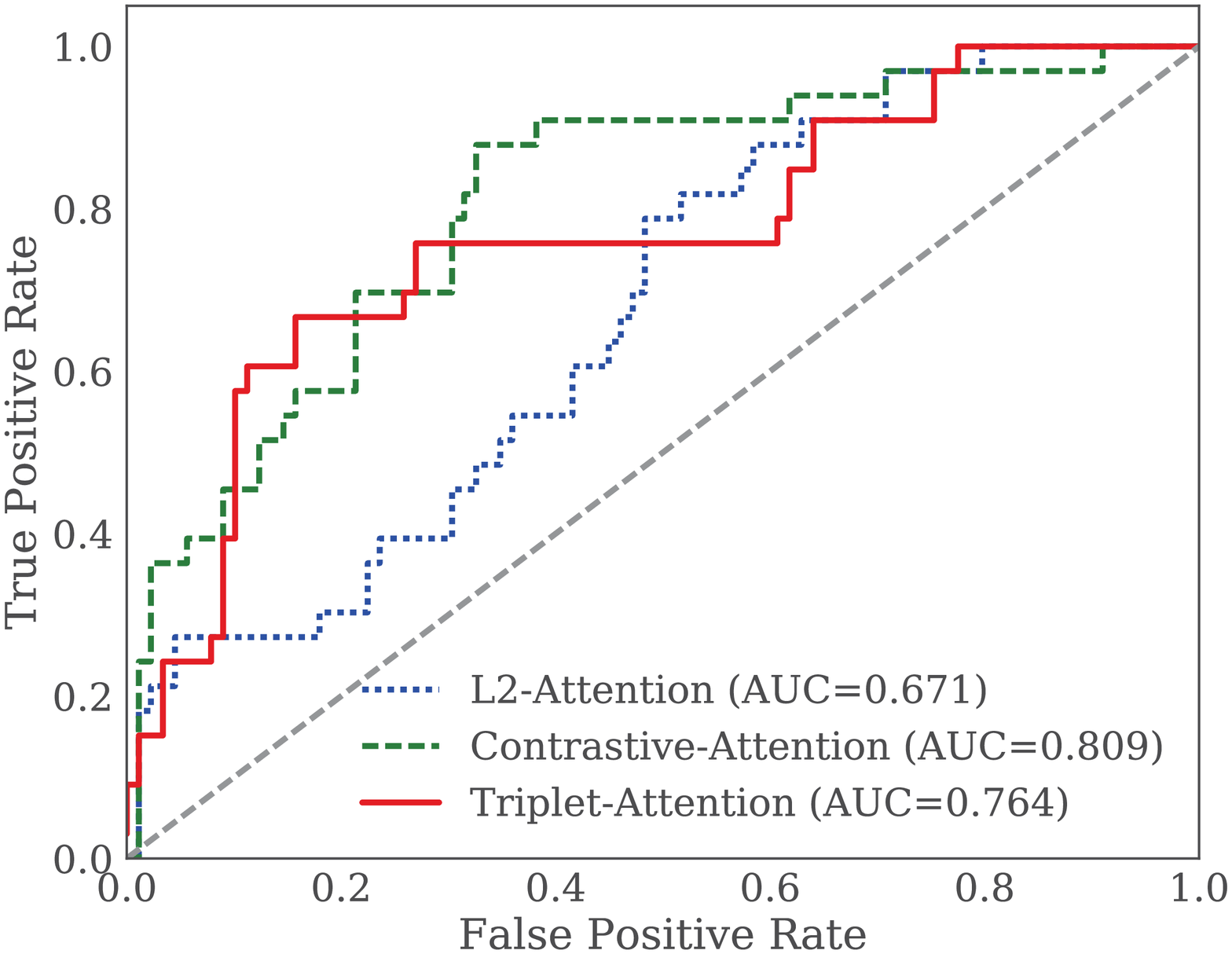}
}
\hfil
\subfigure[]{\includegraphics[width=0.48\columnwidth]{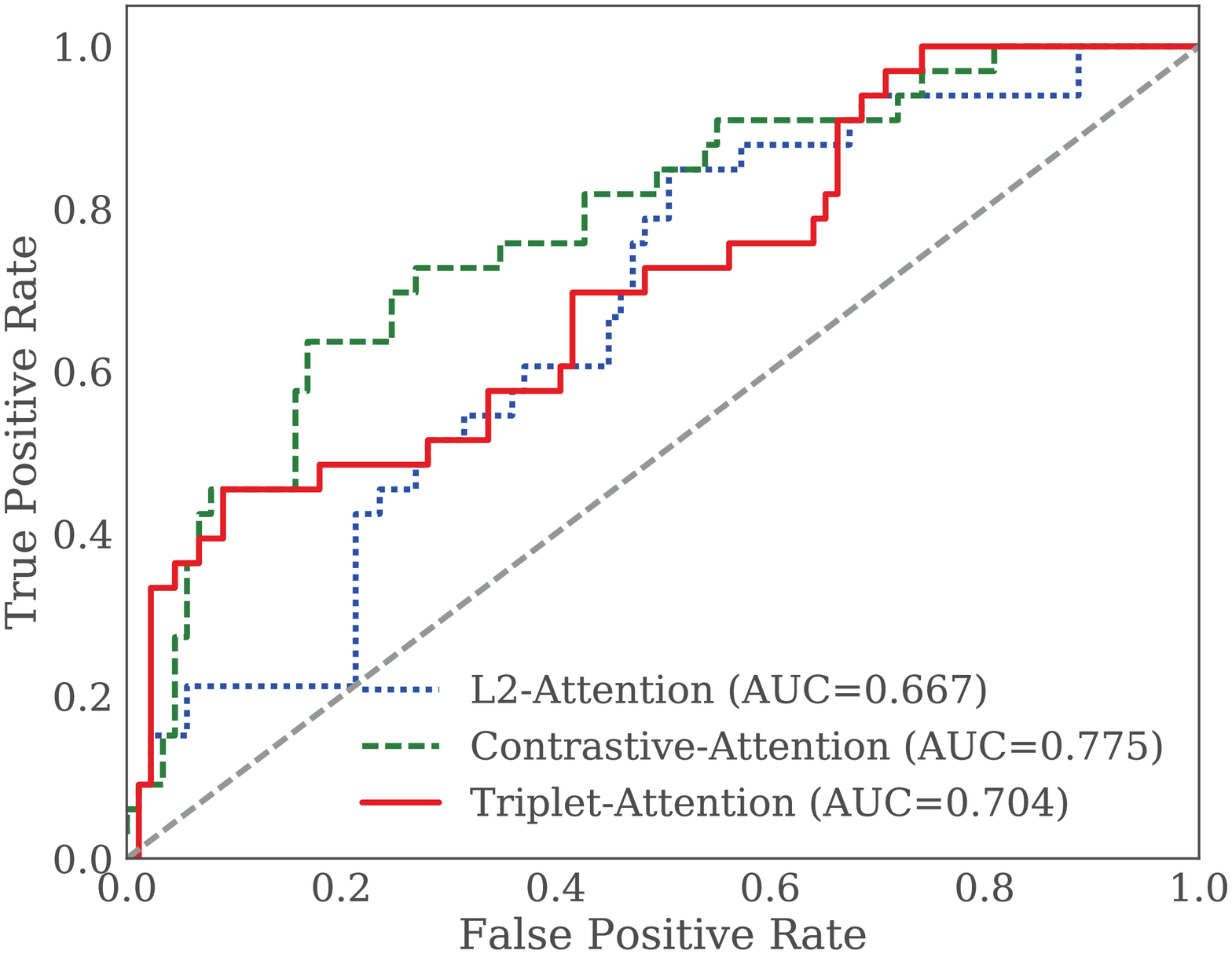}
}
\hfil
\subfigure[]{\includegraphics[width=0.48\columnwidth]{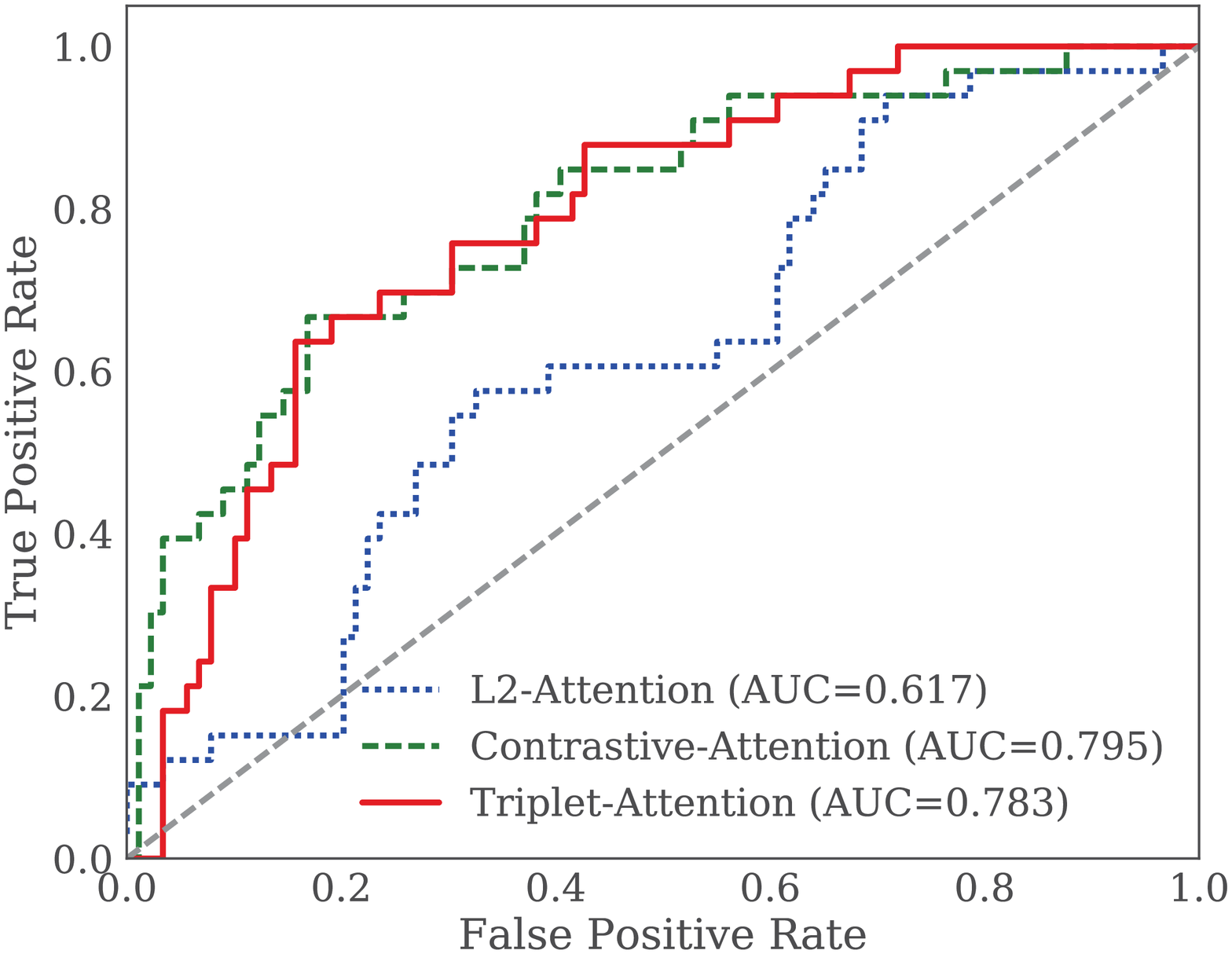}
}
\hfil
\subfigure[]{\includegraphics[width=0.48\columnwidth]{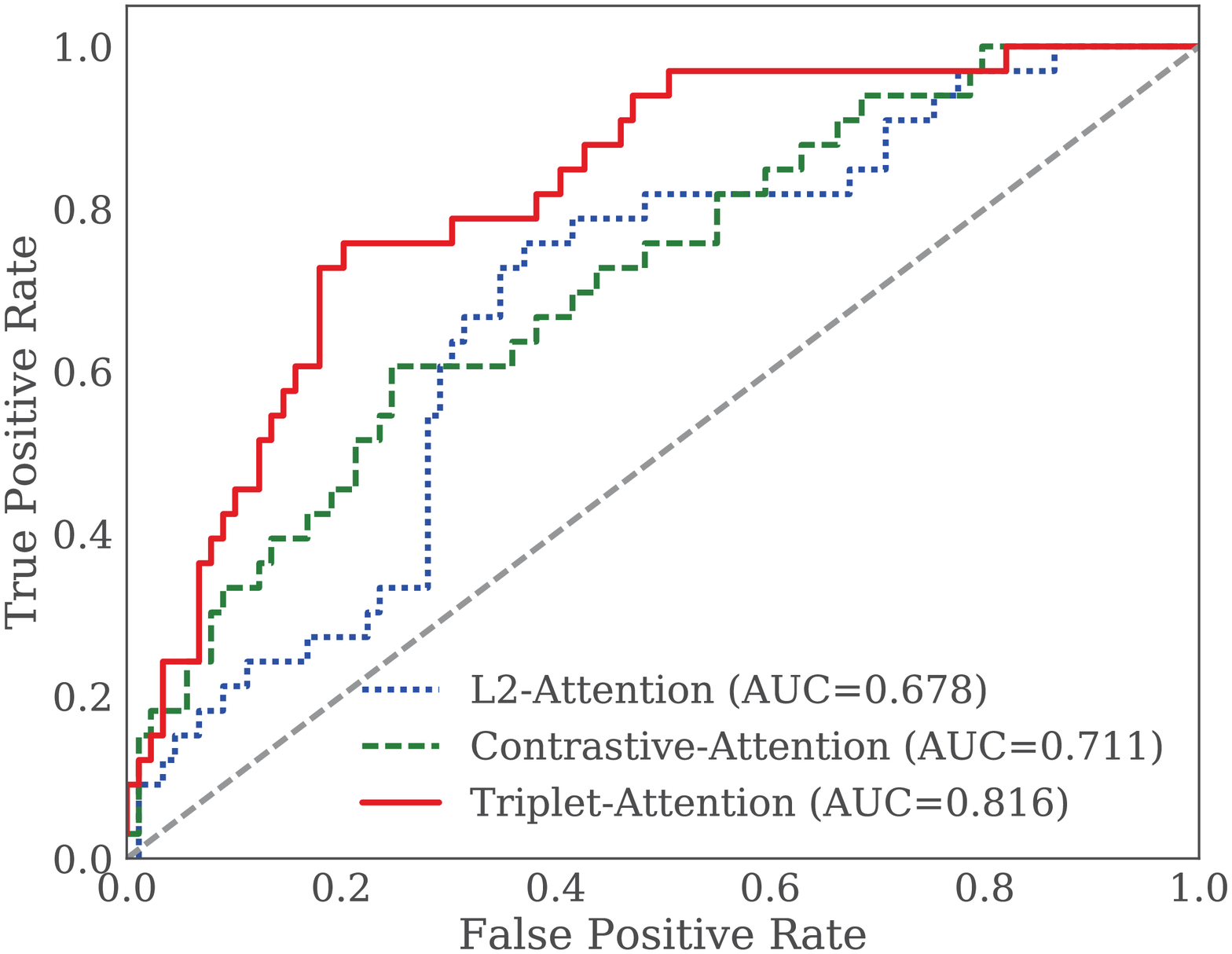}
}
\caption{ROC curves for the $10$ random shuffles of the labeled set (embeddings of multiple skills are weighted averaged using the attention scheme).}
\label{fig:ROC_ATT}
\end{figure*}

\subsection{Experimental Setup}
As presented in Section \ref{sec:Overview}, there are $33$ positive, $89$ negative and $10,221$ unlabeled samples in our dataset. In all the following experiments, we randomly shuffle the $122$ labeled samples and then split the labeled set into five folds, where each fold is used for test in turn. In the remained four folds, three are used for training and one is for validation. We randomly shuffle the labeled set $10$ times to see the stabilities of different models.

\subsubsection{Training Pairs and Triplets Construction}
For the model based on the contrastive loss (Eq. (\ref{eq:Contrast_Loss})), we construct the training pairs using the following strategy: we generate positive pairs by selecting any two samples in the positive or negative set. Suppose we have $N_{+}$ positive and $N_{-}$ negative samples after the split, then the number of positive pairs is $\frac{1}{2}\left( N_{+}(N_{+}-1) + N_{-}(N_{-}-1) \right)$. The negative pairs are generated by selecting one positive and negative sample at each time. This results in $N_{+}N_{-}$ negative pairs.

For the model based on the triplet loss (Eq. (\ref{eq:Triplet_Loss})), the training triplets are constructed by treating the two samples in each of the $\frac{1}{2} \left( N_{+}(N_{+}-1) \right)$ positive pairs as the anchor sample $r_i^a$ and positive sample $r_i^p$ in Eq. (\ref{eq:Triplet_Loss}) respectively. Then each of the $N_{-}$ negative samples are regarded as $r_i^n$, and therefore we have $\frac{1}{2} \left( N_{+}N_{-}(N_{+}-1) \right)$ triplets in total.

For the model that leverages the unlabeled data (Eq. (\ref{eq:MR_Term})), one labeled sample and one of its $k$-nearest neighbor are selected for training at each time. Hence, the number of training pairs is $k(N_{+}+N_{-})$.

\subsubsection{Evaluation Criteria}
Since each fold is used for test in turn, we obtain prediction scores for all the $122$ labeled samples in each random shuffle of the labeled set. We first evaluate the classification performance of different models by using the receiver operating characteristic (ROC) curve, which is able to show the performance under various threshold settings. It depicts relative trade-offs between true positive and false positive. The area under the ROC curve (AUC) is also calculated to illustrate the overall performance. The AUC of a realistic classifier should be larger than $0.5$, and a larger value indicates higher performance. We refer to \cite{T-Fawcett-ML-2004} for a detailed description.

We also map the scores as binary labels using the threshold ``$0$'' and then compare the different models under the F1-measure criterion, which is defined as
\begin{equation}
\notag
F_1 = \frac{2 Prec*Rec}{Prec + Rec},
\end{equation}
where $Prec = \frac{TP}{TP + FP}$ and $Rec = \frac{TP}{TP + FN}$ are the precision and recall respectively. Here, $TP$, $FP$ and $FN$ signify True Positive, False Positive and False Negative respectively.

In addition to classification, we can also sort the different resumes according to the obtained prediction scores. This is helpful when we want to select some resumes from a large corpus. To evaluate the ranking performance of different models, we adopt the popular average precision (AP) criterion:
\begin{equation}
\notag
AP = \frac{\sum_k Prec(k)}{\#\{\mathrm{positive\ samples}\}},
\end{equation}
where $k$ is a rank index of a positive sample and $Prec(k)$ is the precision at the cutoff $k$. It evaluates the fraction of samples ranked above a particular positive sample \cite{M-Zhu-TR-Waterloo-2004}, and a larger value indicates better.

In all the following experiments, we use ``\textbf{L2}'', ``\textbf{Contrastive}'' and ``\textbf{Triplet}'' to denote the baseline model (\ref{eq:L2_Loss}), the model based on contrastive loss (\ref{eq:Contrast_Loss}) and triplet loss (\ref{eq:Triplet_Loss}) respectively. The model that makes use of unlabeled data is denoted as ``\textbf{MR}''.

\subsection{A Comparison of Utilizing Different Losses}
In this set of experiments, we compare the baseline model with its variants of utilizing different losses. The hyper-parameter $\gamma$ in (\ref{eq:General_Objective}) are tuned over the set $\{ 2^i | i = -5, -4, \cdots, 4, 5 \}$. The hidden layer size at the last mapping layers is set as $128$.

We first do not learn the kernel vector $\mathbf{q} \in \mathbb{R}^{512}$ in the different models and directly average the embeddings of multiple skills. This will significantly reduce the number of parameters to be learned and accelerate the training process. The ROC curves of different models for the $10$ random shuffles are shown in Fig.~\ref{fig:ROC_AVG}. From the results, we can see that: 1) in most cases, utilizing either the contrastive or triplet loss significantly outperforms the baseline model, where the least square (L2) loss is adopted. This demonstrates the effectiveness of the variants by changing the loss; 2) Utilizing the triplet loss is usually superior to contrastive loss ($7$ out of $10$) since more training triplets can be generated than training pairs; 3) Sometimes, the model based on contrastive loss is better, see Fig.~\ref{fig:ROC_AVG_3} and Fig.~\ref{fig:ROC_AVG_4}. Besides, in Fig.~\ref{fig:ROC_AVG_6}, we observe that the the AUC score of the baseline model is larger than adopting the contrastive loss. This may be because some noisy pairs or triplets are generated and thus the model is deteriorated.

In Fig.~\ref{fig:ROC_ATT}, we show the ROC curves of different models that employ the attention scheme to weightedly average the embeddings. In general, the results are consistent with those in Fig.~\ref{fig:ROC_AVG}. One main difference is that the two variants always outperform the baseline model. By comparing the results in Fig.~\ref{fig:ROC_ATT_6} with those in Fig.~\ref{fig:ROC_AVG_6}, we can see that when the attention scheme is added, the baseline model becomes worse than utilizing the contrastive loss. This is because when there are more parameters to be learned, the model would be more likely to become over-fitting given the limited training data.

\begin{figure}[htbp]
\centering
\includegraphics[width=0.9\columnwidth]{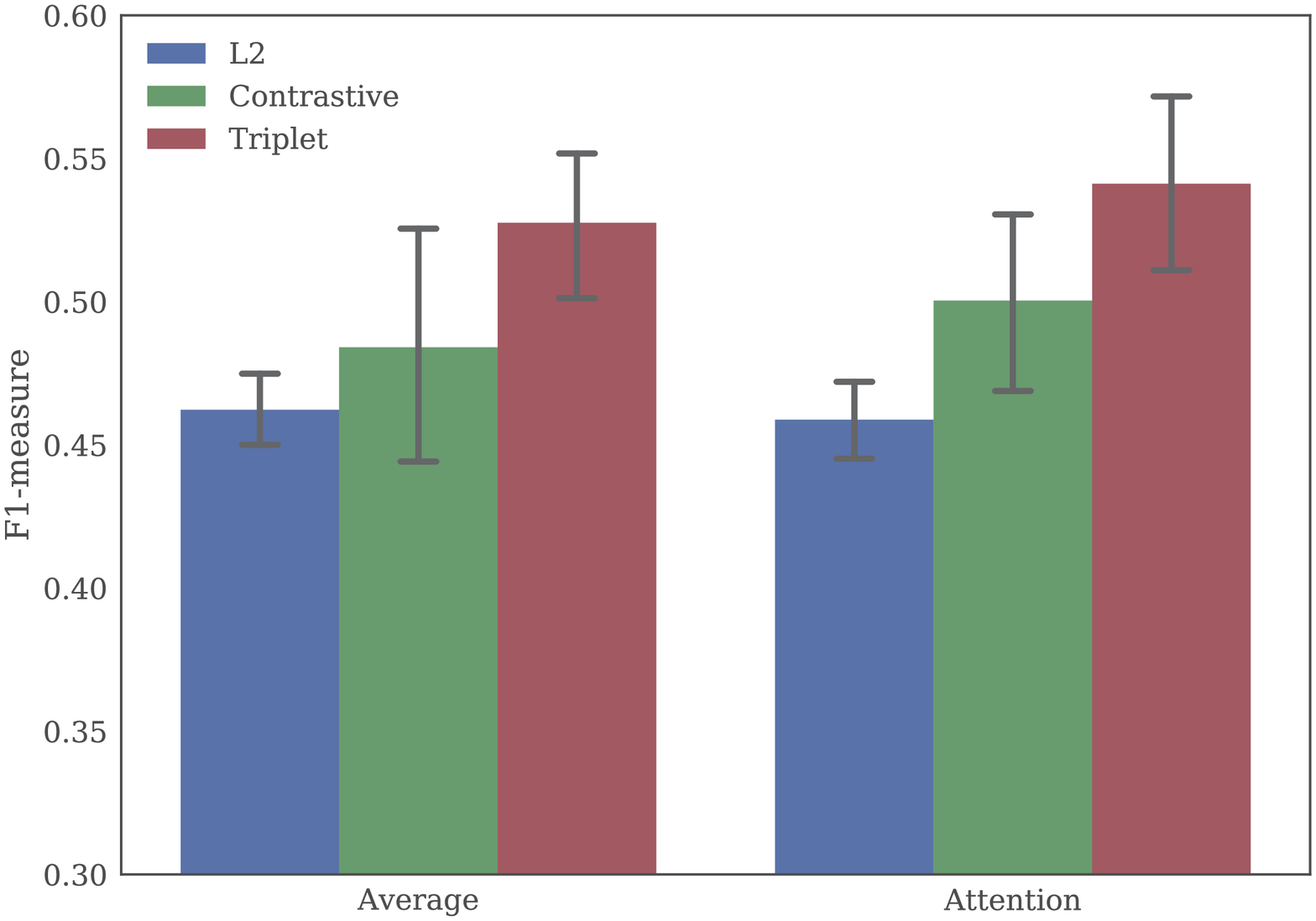}
\caption{F1-measure values of different models (Average: embeddings of multiple skills are simply averaged; Attention: weightedly combine the embeddings of multiple skills using the attention scheme).}
\label{fig:Perf_L_F1}
\end{figure}

\begin{figure}[htbp]
\centering
\includegraphics[width=0.9\columnwidth]{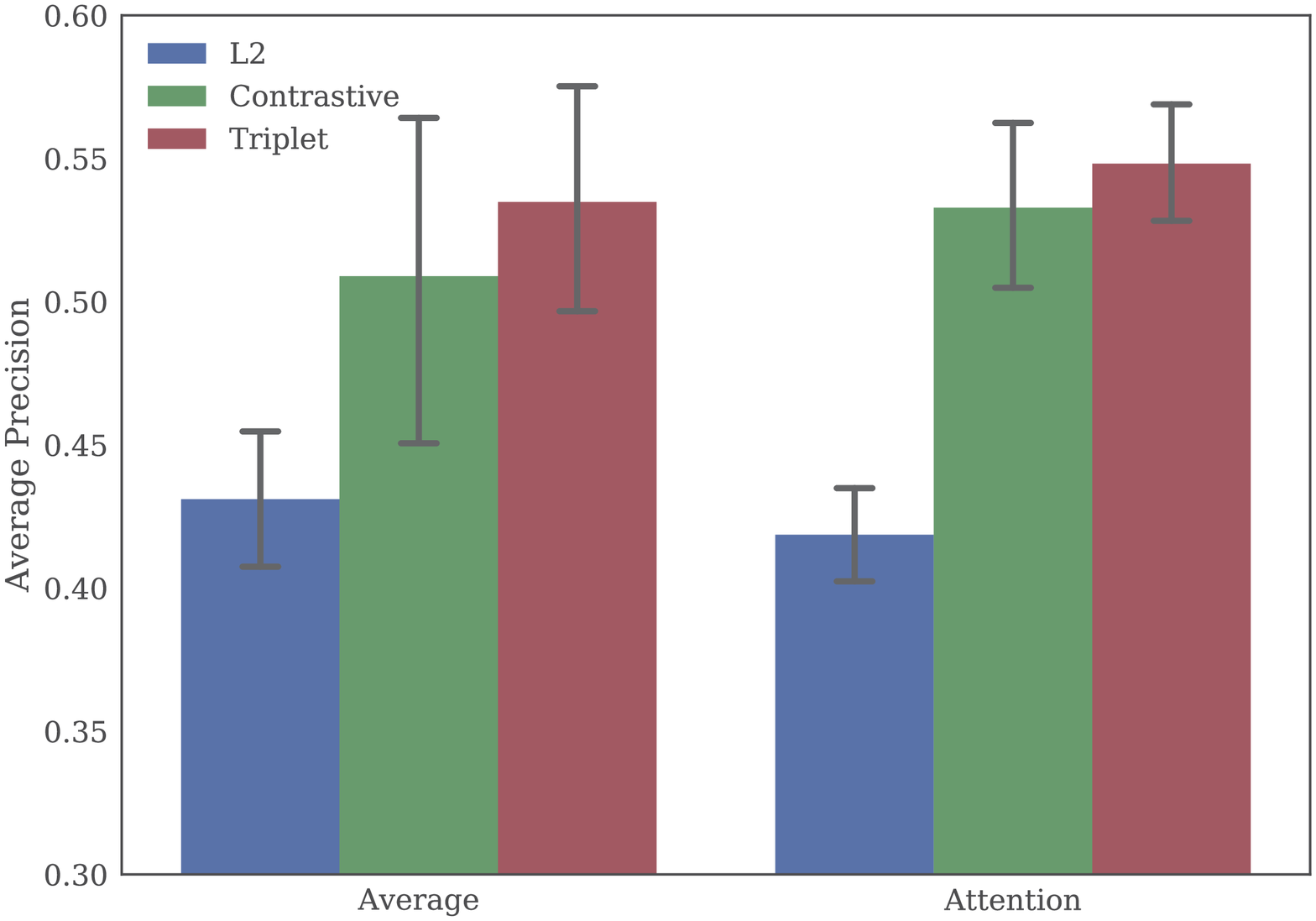}
\caption{Average precisions of different models (Average: embeddings of multiple skills are simply averaged; Attention: weightedly combine the embeddings of multiple skills using the attention scheme).}
\label{fig:Perf_L_AP}
\end{figure}

The F1-measure and AP performance of different models (using the attention scheme or not) are shown in Fig.~\ref{fig:Perf_L_F1} and Fig.~\ref{fig:Perf_L_AP} respectively. From the results, we observe that: utilizing the attention scheme to learn adaptive weights can lead to better performance than the simple average when contrastive or triplet loss is adopted. However, for the baseline model, the performance of simple average are a bit higher. This is also because more parameters are needed to be learned when aggregating with attention. The F1-measure and AP performance are consistent.


\subsection{Making use of Unlabeled Data}
This set of experiments is to evaluate the effectiveness of adding the manifold regularization (MR) term, where the abundant unlabeled data are leveraged. The hyper-parameter $\gamma_I$ in (\ref{eq:MR_Term}) is optimized over the set $\{ 2^i | i = -8, -4, \cdots, 4, 2 \}$ and the number of nearest neighbors $k$ is chosen from $\{5, 10, 15, 20, 30\}$.

\begin{figure}[htbp]
\centering
\includegraphics[width=0.9\columnwidth]{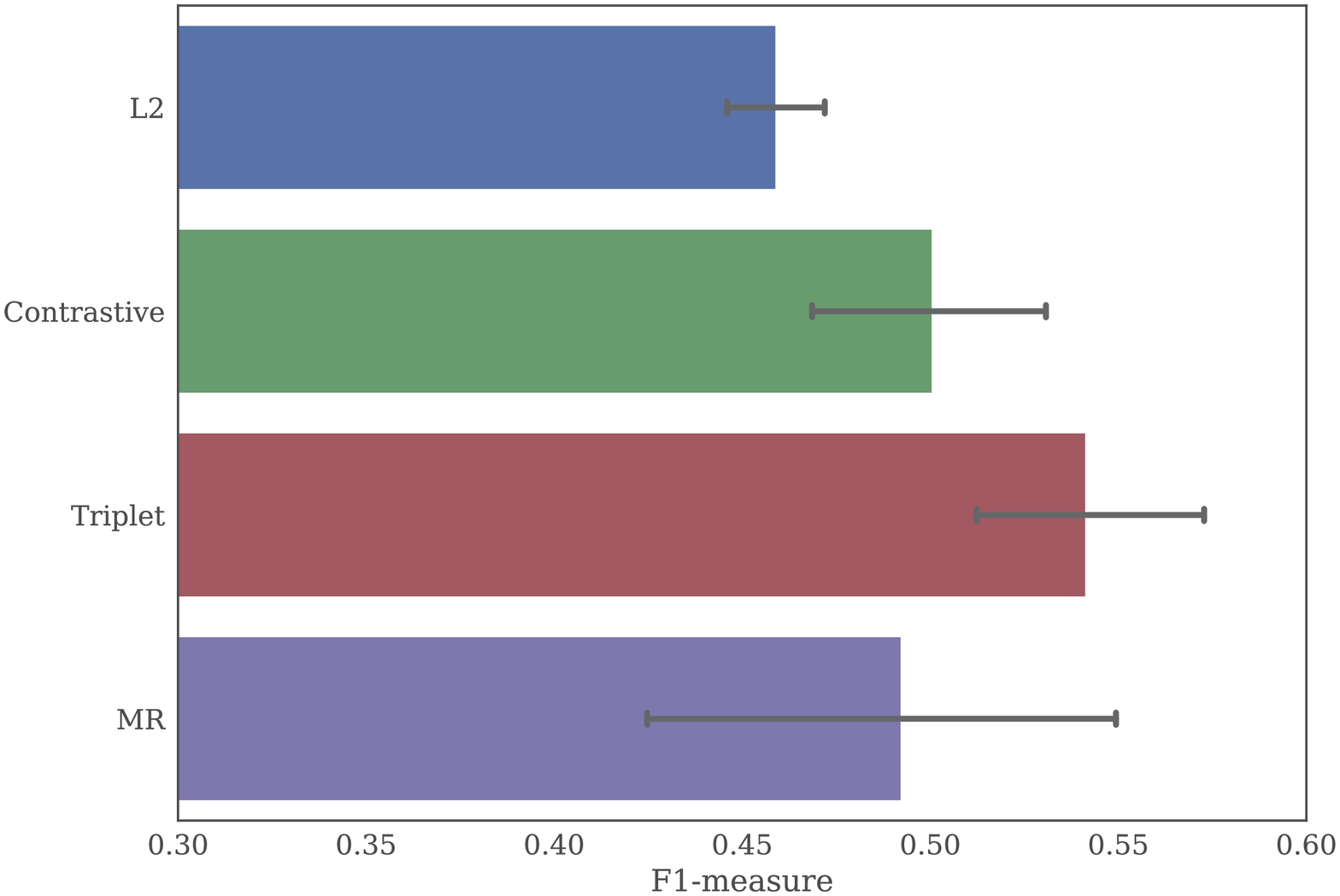}
\caption{Comparison of utilizing different losses with MR in terms of F1-measure.}
\label{fig:Perf_U_F1}
\end{figure}

\begin{figure}[htbp]
\centering
\includegraphics[width=0.9\columnwidth]{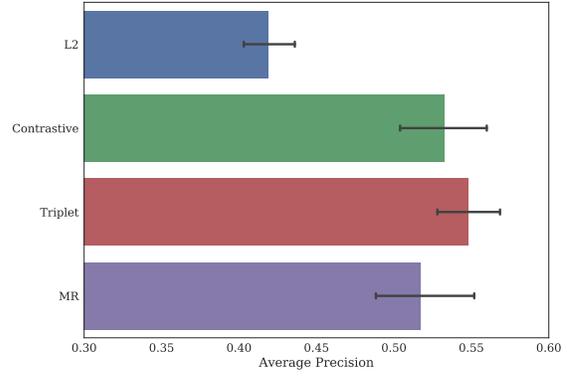}
\caption{Comparison of utilizing different losses with MR in terms of average precision.}
\label{fig:Perf_U_AP}
\end{figure}

The F1-measure and AP performance of different models using the attention scheme are shown in Fig.~\ref{fig:Perf_U_F1} and Fig.~\ref{fig:Perf_U_AP} respectively. From the results, we can see that adding the manifold regularization term is helpful to alleviate the label deficiency issue, but it is not more advantageous than constructing training pairs or triplets using the limited labeled data.


\begin{figure}[htbp]
\centering
\includegraphics[width=0.9\columnwidth]{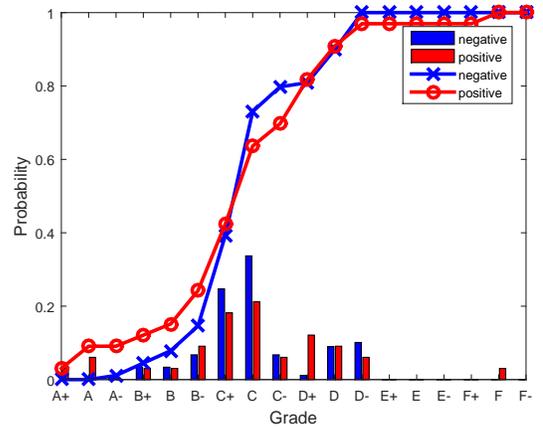}
\caption{Statistics of the positive and negative samples that assigned to different grades. The bars are probabilities and the curves are accumulative sums of the probabilities.}
\label{fig:Rezscore_StatCum}
\end{figure}

\subsection{A Comparison with Other Approach}
Finally, we compare the proposed framework with other approaches. Existing quality assessment algorithms are usually designed for visual contents \cite{WW-Zhang-et-al-TMM-2013, J-He-et-al-TCSVT-2014, KD-Ma-et-al-TIP-2017, GY-Gao-et-al-TMM-2018, KD-Ma-et-al-TIP-2018, HZ-Zhang-et-al-arXiv-2018}, and not appropriate for the textual resumes studied in this paper. Since there is no public algorithm for RQA, we submit our labeled resumes to a website (http://rezscore.com/), which can assign a grade for each resume. There are $18$ grades: $\mathrm{A+}, \mathrm{A}, \mathrm{A-}, \cdots, \mathrm{F+}, \mathrm{F}, \mathrm{F-}$. The distributions over the different grades for the positive and negative samples are shown in Fig.~\ref{fig:Rezscore_StatCum}, where the curves are accumulative sums of the probabilities.

It can be seen from the figure that the positive probabilities are larger than negative ones when the grade is $\mathrm{C+}$ or higher. Therefore, we choose $\mathrm{C+}$ as the threshold. That is, the samples that assigned with a grade $\mathrm{C+}$ or higher is predicted as positive samples and otherwise negative. We then calculate the F1-measure values and compare them with those of our different models. The performance in terms of ROC curve and AP are not available since the outputs are not continuous-valued scores. The compared approach is denoted as ``\textbf{Rezscore}'' and the results are reported in Table~\ref{tab:Other_F1}. It can be observed from the results that Rezscore is even worse than the proposed baseline model. This further demonstrates superiority of the proposed framework for RQA.

\begin{table}[htbp]
\caption{A comparison of our models with the other approach in terms of F1-measure.}
\centering
\begin{tabular}{|c|c|}
\hline
Methods & F1-measure \\
\hline
L2 & 0.459$\pm$0.022 \\
Contrastive & 0.500$\pm$0.054 \\
Triplet & 0.541$\pm$0.051 \\
\hline
MR & 0.492$\pm$0.109 \\
\hline
Rezscore & 0.341 \\
\hline
\end{tabular}
\label{tab:Other_F1}
\end{table}

\section{Conclusion}
\label{sec:Conclusion}
In this paper, we study the problem of automatic resume quality assessment (RQA), which is essential but has received little attention. We build a dataset, identify some useful features/factors and propose a general framework for RQA. To deal with the common issue of limited labeled data, we tried several different strategies, such as adopting the pair/triplet-based loss and making use of unlabeled data. From the results, we mainly conclude that: 1) learning adaptive weights using the attention scheme to aggregate multiple embeddings is superior to the simple average in general. This is consistent with the literature results of aggregation with attention, but worse results may be obtained when the training data is insufficient; 2) either using the designed pair/triplet-based loss or adding a regularization term to utilize unlabeled data can improve the performance, it seems that the model based on triplet loss achieves the best performance overall. In the future, we plan to combine the different strategies to further improve the performance, build a larger corpus that includes job-post information and identify more useful features for RQA. Besides, we anticipate that automatic RQA would contribute significantly to transform the future of human resources management.




%

\section*{Acknowledgment}
This work is supported by Singapore NRF2015ENCGDCR01001-003, administrated via IMDA and NRF2015ENCGBICRD001-012, administrated via BCA.

\bibliographystyle{IEEEtran}
\bibliography{IEEEabrv,./ICDM-18_paper_DM526}

\begin{thebibliography}{10}
\providecommand{\url}[1]{#1}
\csname url@samestyle\endcsname
\providecommand{\newblock}{\relax}
\providecommand{\bibinfo}[2]{#2}
\providecommand{\BIBentrySTDinterwordspacing}{\spaceskip=0pt\relax}
\providecommand{\BIBentryALTinterwordstretchfactor}{4}
\providecommand{\BIBentryALTinterwordspacing}{\spaceskip=\fontdimen2\font plus
\BIBentryALTinterwordstretchfactor\fontdimen3\font minus
  \fontdimen4\font\relax}
\providecommand{\BIBforeignlanguage}[2]{{%
\expandafter\ifx\csname l@#1\endcsname\relax
\typeout{** WARNING: IEEEtran.bst: No hyphenation pattern has been}%
\typeout{** loaded for the language `#1'. Using the pattern for}%
\typeout{** the default language instead.}%
\else
\language=\csname l@#1\endcsname
\fi
#2}}
\providecommand{\BIBdecl}{\relax}
\BIBdecl

\bibitem{D-Cer-et-al-arXiv-2018}
D.~Cer, Y.~Yang, S.-y. Kong, N.~Hua, N.~Limtiaco, R.~S. John, N.~Constant,
  M.~Guajardo-Cespedes, S.~Yuan, C.~Tar \emph{et~al.}, ``Universal sentence
  encoder,'' \emph{arXiv preprint arXiv:1803.11175}, 2018.

\bibitem{JL-Yang-et-al-CVPR-2017}
J.~Yang, P.~Ren, D.~Zhang, D.~Chen, F.~Wen, H.~Li, and G.~Hua, ``Neural
  aggregation network for video face recognition,'' in \emph{IEEE Conference on
  Computer Vision and Pattern Recognition}, 2017, pp. 4362--4371.

\bibitem{X-Zhu-TR-Madison-2006}
X.~Zhu, ``Semi-supervised learning literature survey,'' Dept. Electr. Eng.,
  Univ. Wisconsin-Madison, Madison, Tech. Rep. 1530, 2006.

\bibitem{O-Chapelle-et-al-TNN-2009}
O.~Chapelle, B.~Scholkopf, and A.~Zien, ``Semi-supervised learning (chapelle,
  o. et al., eds.; 2006)[book reviews],'' \emph{IEEE Transactions on Neural
  Networks}, vol.~20, no.~3, pp. 542--542, 2009.

\bibitem{A-Isabelle-et-al-NIPSw-2017}
I.~Augenstein, S.~Bach, E.~Belilovsky, M.~Blaschko, C.~Lampert, E.~Oyallon,
  E.~A. Platanios, A.~Ratner, and C.~Re, ``Learning with limited labeled data:
  Weak supervision and beyond,'' in \emph{Advances in Neural Information
  Processing Systems Workshop}, 2017.

\bibitem{N-Japkowicz-and-S-Stephen-IDA-2002}
N.~Japkowicz and S.~Stephen, ``The class imbalance problem: A systematic
  study,'' \emph{Intelligent Data Analysis}, vol.~6, no.~5, pp. 429--449, 2002.

\bibitem{S-Chopra-et-al-CVPR-2005}
S.~Chopra, R.~Hadsell, and Y.~LeCun, ``Learning a similarity metric
  discriminatively, with application to face verification,'' in \emph{IEEE
  Conference on Computer Vision and Pattern Recognition}, 2005, pp. 539--546.

\bibitem{M-Norouzi-et-al-NIPS-2012}
M.~Norouzi, D.~J. Fleet, and R.~R. Salakhutdinov, ``Hamming distance metric
  learning,'' in \emph{Advances in Neural Information Processing Systems},
  2012, pp. 1061--1069.

\bibitem{F-Schroff-et-al-CVPR-2015}
F.~Schroff, D.~Kalenichenko, and J.~Philbin, ``Facenet: A unified embedding for
  face recognition and clustering,'' in \emph{IEEE Conference on Computer
  Vision and Pattern Recognition}, 2015, pp. 815--823.

\bibitem{M-Belkin-et-al-JMLR-2006}
M.~Belkin, P.~Niyogi, and V.~Sindhwani, ``Manifold regularization: A geometric
  framework for learning from labeled and unlabeled examples,'' \emph{Journal
  of Machine Learning Research}, vol.~7, pp. 2399--2434, 2006.

\bibitem{H-Gui-et-al-ICDM-2016}
H.~Gui, J.~Liu, F.~Tao, M.~Jiang, B.~Norick, and J.~Han, ``Large-scale
  embedding learning in heterogeneous event data,'' in \emph{International
  Conference on Data Mining}, 2016, pp. 907--912.

\bibitem{Y-Bengio-et-al-JMLR-2003}
Y.~Bengio, R.~Ducharme, P.~Vincent, and C.~Jauvin, ``A neural probabilistic
  language model,'' \emph{Journal of Machine Learning Research}, vol.~3, pp.
  1137--1155, 2003.

\bibitem{GX-Xun-et-al-ICDM-2016}
G.~Xun, V.~Gopalakrishnan, F.~Ma, Y.~Li, J.~Gao, and A.~Zhang, ``Topic
  discovery for short texts using word embeddings,'' in \emph{International
  Conference on Data Mining}, 2016, pp. 1299--1304.

\bibitem{J-Turian-et-al-ACL-2010}
J.~Turian, L.~Ratinov, and Y.~Bengio, ``Word representations: a simple and
  general method for semi-supervised learning,'' in \emph{Annual Meeting of the
  Association for Computational Linguistics}, 2010, pp. 384--394.

\bibitem{T-Mikolov-et-al-NIPS-2013}
T.~Mikolov, I.~Sutskever, K.~Chen, G.~S. Corrado, and J.~Dean, ``Distributed
  representations of words and phrases and their compositionality,'' in
  \emph{Advances in Neural Information Processing Systems}, 2013, pp.
  3111--3119.

\bibitem{J-Pennington-et-al-EMNLP-2014}
J.~Pennington, R.~Socher, and C.~Manning, ``Glove: Global vectors for word
  representation,'' in \emph{Proceedings of the Conference on Empirical Methods
  in Natural Language Processing}, 2014, pp. 1532--1543.

\bibitem{A-Conneau-et-al-EMNNLP-2017}
A.~Conneau, D.~Kiela, H.~Schwenk, L.~Barrault, and A.~Bordes, ``Supervised
  learning of universal sentence representations from natural language
  inference data,'' in \emph{Proceedings of the Conference on Empirical Methods
  in Natural Language Processing}, 2017, pp. 670--680.

\bibitem{B-Graham-arXiv-2014}
B.~Graham, ``Fractional max-pooling,'' \emph{arXiv preprint arXiv:1412.6071},
  2014.

\bibitem{Y-Luo-et-al-TKDE-2015}
Y.~Luo, D.~Tao, K.~Ramamohanarao, C.~Xu, and Y.~Wen, ``Tensor canonical
  correlation analysis for multi-view dimension reduction,'' \emph{IEEE
  transactions on Knowledge and Data Engineering}, vol.~27, no.~11, pp.
  3111--3124, 2015.

\bibitem{Y-Luo-et-al-TIP-2015}
Y.~Luo, T.~Liu, D.~Tao, and C.~Xu, ``Multiview matrix completion for multilabel
  image classification,'' \emph{IEEE Transactions on Image Processing},
  vol.~24, no.~8, pp. 2355--2368, 2015.

\bibitem{YJ-Fu-et-al-ICDM-2016}
Y.~Fu, J.~Liu, X.~Li, X.~Lu, J.~Ming, C.~Guan, and H.~Xiong, ``Service usage
  analysis in mobile messaging apps: A multi-label multi-view perspective,'' in
  \emph{International Conference on Data Mining}, 2016, pp. 877--882.

\bibitem{O-Vinyals-et-al-ICLR-2016}
O.~Vinyals, S.~Bengio, and M.~Kudlur, ``Order matters: Sequence to sequence for
  sets,'' in \emph{International Conference on Learning Representation}, 2016.

\bibitem{S-Poriaa-et-al-ICDM-2017}
S.~Poriaa, E.~Cambriab, D.~Hazarikac, N.~Mazumderd, A.~Zadehe, and L.-P.
  Morencye, ``Multi-level multiple attentions for contextual multimodal
  sentiment analysis,'' in \emph{International Conference on Data Mining},
  2017, pp. 1033--1038.

\bibitem{Y-Luo-et-al-TIP-2013}
Y.~Luo, D.~Tao, B.~Geng, C.~Xu, and S.~J. Maybank, ``Manifold regularized
  multitask learning for semi-supervised multilabel image classification,''
  \emph{IEEE Transactions on Image Processing}, vol.~22, no.~2, pp. 523--536,
  2013.

\bibitem{Y-Luo-et-al-TIP-2014}
Y.~Luo, T.~Liu, D.~Tao, and C.~Xu, ``Decomposition-based transfer distance
  metric learning for image classification.'' \emph{IEEE Transactions on Image
  Processing}, vol.~23, no.~9, pp. 3789--3801, 2014.

\bibitem{Y-Luo-et-al-TPAMI-2018}
Y.~Luo, Y.~Wen, T.~Liu, and D.~Tao, ``Transferring knowledge fragments for
  learning distance metric from a heterogeneous domain,'' \emph{IEEE
  Transactions on Pattern Analysis and Machine Intelligence}, 2018.

\bibitem{Y-Luo-et-al-TNNLS-2013}
Y.~Luo, D.~Tao, C.~Xu, C.~Xu, H.~Liu, and Y.~Wen, ``Multiview vector-valued
  manifold regularization for multilabel image classification,'' \emph{IEEE
  Transactions on Neural Networks and Learning Systems}, vol.~24, no.~5, pp.
  709--722, 2013.

\bibitem{D-Cheng-et-al-CVPR-2016}
D.~Cheng, Y.~Gong, S.~Zhou, J.~Wang, and N.~Zheng, ``Person re-identification
  by multi-channel parts-based cnn with improved triplet loss function,'' in
  \emph{IEEE Conference on Computer Vision and Pattern Recognition}, 2016, pp.
  1335--1344.

\bibitem{J-Wang-et-al-CVPR-2014}
J.~Wang, Y.~Song, T.~Leung, C.~Rosenberg, J.~Wang, J.~Philbin, B.~Chen, and
  Y.~Wu, ``Learning fine-grained image similarity with deep ranking,'' in
  \emph{IEEE Conference on Computer Vision and Pattern Recognition}, 2014, pp.
  1386--1393.

\bibitem{T-Fawcett-ML-2004}
T.~Fawcett, ``{ROC} graphs: Notes and practical considerations for
  researchers,'' \emph{Machine Learning}, vol.~31, pp. 1--38, 2004.

\bibitem{M-Zhu-TR-Waterloo-2004}
M.~Zhu, ``Recall, precision and average precision,'' Dept. Electr. Eng., Univ.
  Waterloo, Tech. Rep. Working Paper 2004-09, 2004.

\bibitem{WW-Zhang-et-al-TMM-2013}
W.~Zhang, Y.~Wen, Z.~Chen, and A.~Khisti, ``{QoE}-driven cache management for
  http adaptive bit rate streaming over wireless networks,'' \emph{IEEE
  Transactions on Multimedia}, vol.~15, no.~6, pp. 1431--1445, 2013.

\bibitem{J-He-et-al-TCSVT-2014}
J.~He, Y.~Wen, J.~Huang, and D.~Wu, ``On the {Cost-QoE} tradeoff for
  cloud-based video streaming under amazon {EC2}'s pricing models,'' \emph{IEEE
  Transactions on Circuits and Systems for Video Technology}, vol.~24, no.~4,
  pp. 669--680, 2014.

\bibitem{KD-Ma-et-al-TIP-2017}
K.~Ma, W.~Liu, T.~Liu, Z.~Wang, and D.~Tao, ``{dipIQ}: Blind image quality
  assessment by learning-to-rank discriminable image pairs,'' \emph{IEEE
  Transactions on Image Processing}, vol.~26, no.~8, pp. 3951--3964, 2017.

\bibitem{GY-Gao-et-al-TMM-2018}
G.~Gao, H.~Zhang, H.~Hu, Y.~Wen, J.~Cai, C.~Luo, and W.~Zeng, ``Optimizing
  quality of experience for adaptive bitrate streaming via viewer interest
  inference,'' \emph{IEEE Transactions on Multimedia}, 2018.

\bibitem{KD-Ma-et-al-TIP-2018}
K.~Ma, W.~Liu, K.~Zhang, Z.~Duanmu, Z.~Wang, and W.~Zuo, ``End-to-end blind
  image quality assessment using deep neural networks,'' \emph{IEEE
  Transactions on Image Processing}, vol.~27, no.~3, pp. 1202--1213, 2018.

\bibitem{HZ-Zhang-et-al-arXiv-2018}
H.~Zhang, H.~Hu, G.~Gao, Y.~Wen, and K.~Guan, ``{DeepQoE}: A unified framework
  for learning to predict video {QoE},'' \emph{arXiv preprint
  arXiv:1804.03481}, 2018.

\end{thebibliography}

\end{document}